\documentclass[final,superscriptaddress,english,twocolumn,amssymb, nobibnotes, aps, prb, longbibliography]{revtex4-1}
\usepackage{graphicx}
\usepackage{natbib}
\newcommand{\tr}{\mathrm{Tr}}
\usepackage{amsmath}
\usepackage{amssymb}
\usepackage{appendix}
\usepackage[dvipsnames]{xcolor}{\huge }
\usepackage[section]{placeins}
\definecolor{darkblue}{rgb}{0,0,.65}
\definecolor{darkgreen}{rgb}{0.28,0.41,0.19}

\usepackage{wrapfig}{\LARGE }
\usepackage{nicefrac}
\usepackage[pdfauthor={David J. Luitz},pdfstartview=FitH,breaklinks=true,bookmarks=true,colorlinks=true,anchorcolor=black,citecolor=blue,
  filecolor=black,menucolor=black,urlcolor=darkblue,linkcolor=blue,]{hyperref}
\usepackage[all]{hypcap} \newcommand{\bra}[1]{\langle\,#1\,|}
\newcommand{\ket}[1]{|#1\rangle}

\newcommand{\braket}[2]{\langle\,#1\, | \, #2\,\rangle}

\newcommand{\mat}[1]{\mathsf{#1}}

\newcommand{\E}{\mathrm{e}}
\newcommand{\I}{\mathrm{i}}
\newcommand{\D}{\mathrm{d}}

\newcommand{\up}{\uparrow}

\newcommand{\eye}{\mathsf{1}}

\graphicspath{{images/}}
\begin{document}

\title{The pyrochlore $S=1/2$ Heisenberg antiferromagnet at finite temperature}

\author{Robin Sch\"afer}
\affiliation{Max Planck Institute for the Physics of Complex Systems, Noethnitzer Str. 38, 01187 Dresden, Germany}
\author{Imre Hagym\'asi}
\affiliation{Max Planck Institute for the Physics of Complex Systems, Noethnitzer Str. 38, 01187 Dresden, Germany}
\affiliation{Strongly Correlated Systems "Lend\"ulet" Research Group, Institute 
for Solid State
Physics and Optics, Wigner Research Centre for Physics, Budapest H-1525 P.O. 
Box 49, Hungary
}
\author{Roderich Moessner}
\author{David J. Luitz}
\email{dluitz@pks.mpg.de}
\affiliation{Max Planck Institute for the Physics of Complex Systems, Noethnitzer Str. 38, 01187 Dresden, Germany}

\date{\today}

\begin{abstract}
Frustrated three dimensional quantum magnets are notoriously impervious to theoretical analysis. 
Here we use a combination of 
three computational methods to investigate the three dimensional pyrochlore $S=1/2$ quantum antiferromagnet, an archetypical frustrated magnet, at finite temperature, $T$: canonical typicality for a finite cluster of $2\times 2 \times 2$ unit cells (i.e. $32$ sites), a finite-$T$ matrix product state method on a larger cluster with $48$ sites, and the numerical linked cluster expansion (NLCE) using clusters up to $25$ lattice sites, which include non-trivial hexagonal and octagonal loops. We focus on thermodynamic
properties (energy, specific heat capacity, entropy, susceptibility, magnetisation) next to the static structure factor. 
We find a pronounced maximum in the specific heat at $T = 0.57 J$, which is stable across finite size clusters and converged in the series expansion. This is well-separated from a residual amount of spectral weight of $0.47 k_B \ln2$ per spin which
has not been released even at $T\approx0.25 J$, the limit of convergence of our results. This is a large
value compared to a number of highly frustrated models and materials, such as spin ice or the kagome $S=1/2$ Heisenberg
antiferromagnet. 
We also find a non-monotonic
dependence on $T$ of the magnetisation at low magnetic fields, reflecting the dominantly non-magnetic character of the low-energy spectral weight. 
A detailed comparison of our results to measurements for the $S=1$ material NaCaNi$_2$F$_7$ yields rough agreement of the functional form of the specific heat maximum, which in turn differs from the sharper maximum 
of the heat capacity of the spin ice material Dy$_2$Ti$_2$O$_7$, all of which are yet qualitatively distinct from conventional,
unfrustrated magnets.
\end{abstract}
\maketitle

\section{Introduction}
The pyrochlore lattice, composed of corner-sharing tetrahedra, is a common motif in 
materials chemistry; in the context of magnetic materials, it has been prominent
in a range of rare-earth\cite{ggg_review}  and spinel compounds\cite{bragg_structure_1915,Spinel_review_jpsj}.
Pyrochlore magnets and models have played a tremendously important role in the history of 
frustrated magnetism and topological condensed matter physics. One of the foundational 
publications, in 1956, 
was Anderson's identification of the classical pyrochlore Ising magnet \cite{anderson_ferrite}
as an interesting model system. Now called spin ice\cite{harbram}, this is a topological magnet
exhibiting an emergent gauge field and fractionalised excitations \cite{castelnovo_spin_2012}.

\begin{figure}[h]
    \centering
    \includegraphics{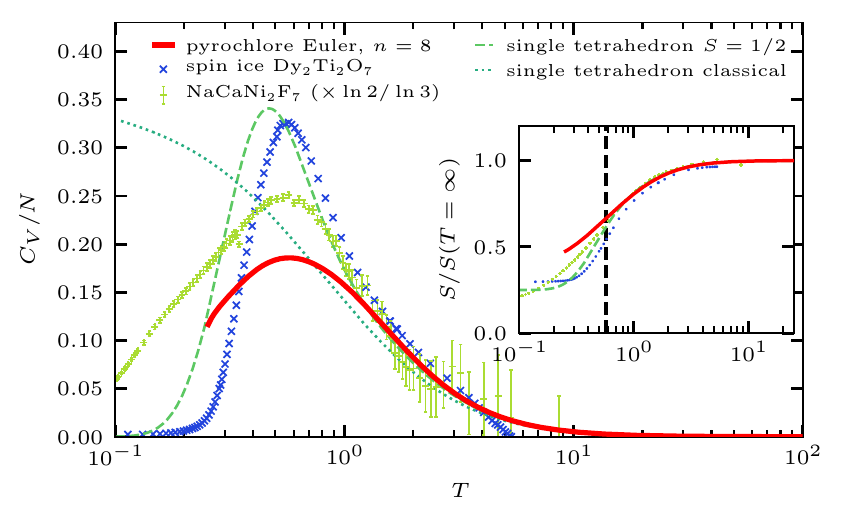}
    \caption{Heat capacity per spin in different pyrochlore magnets, for a detailed account see Sec.~\ref{sec:results}. The red curve represents the converged part of our results from the NLCE for the  $S=1/2$ Heisenberg model. Results for 
   a single tetrahedron for  $S=1/2$ (dashed) for $S=\infty$ (classical case, dots) have $T$ scaled to match in the high-$T$ limit. Symbols are for experiments on NaCaNi$_2$F$_7$\cite{plumb_continuum_2019},  a $S=1$ (approximate) Heisenberg magnet, and the Ising spin ice Dy$_2$Ti$_2$O$_7$\cite{ramirez_zero-point_1999}, both scaled in $T$ such that their maxima coincide with that of the $S=1/2$ model (with a factor $\ln 3/\ln 2$ to account for the larger $S=1$ entropy). Inset similarly shows entropy per spin.
   }
   \label{fig:specheat_exp}
\end{figure}

The classical Heisenberg model, following a pioneering study by Villain\cite{villain_insulating_1979}, turned out
to be the first classical Heisenberg spin liquid\cite{moecha_pyro_prl}. This undergoes a {\it very} delicate
order-by-disorder transition for large spins, as the zero-point energy induced by 
quantum fluctuations favours a subset
of collinear states\cite{hizi_anharmonic_2009,Henley_pyro_largeS}.
Beyond this (semi-)classical 
limit of large spin, $S\rightarrow\infty$, little is known reliably about the properties
of the pyrochlore quantum Heisenberg model. 

This is because the properties of the lattice
conspire to frustrate not only magnetic order, but also 
attempts to apply standard theoretical and numerical approaches. 
The presence of a macroscopic
number (`flat band') of gapless excitations  in most bare models precludes standard 
perturbative schemes and mean-field theories.  This reflects the fact that fluctuations
are typically very strong, the basic ingredient via which frustrated magnets avoid ordering.
For this reason, even the relatively 'high' dimensionality, $d=3$, often considered almost homologous
with proximity to mean-field behaviour, is a hindrance rather than
a help: the most unbiased method, exact diagonalisation, breaks down already for small linear
system sizes $L$, as the Hilbert space dimension grows exponentially with $L^3$. 
Similarly, DMRG-based methods are well-known to struggle beyond $d=1$, 
while geometric frustration yields a sign problem in quantum Monte Carlo. 
In this sense, the pyrochlore magnet is even less tractable than the notoriously enigmatic 
kagome $S=1/2$ Heisenberg magnet, for which decades
of intensive interest have not yielded a consensus on the nature of its ground state.
For the pyrochlore lattice, even a reliable 
ground-state energy estimate is lacking, and the proposed ground 
states depend strongly on the method used to study it.\cite{harris_ordering_1998,isoda_valence_bond_1998,CanalsLacroix_prl,Berg_subcontractor_2003,MoessnerGoerbig_prb_2006,burnell_monopole_2009,Nussinov_copy,tsunetsugu_theory_2017,chandra_spin12_2018,GarciaHuber_qtet_prb,iqbal_quantum_2019} 

The pyrochlore $S=1/2$ Heisenberg magnet thus has many ingredients 
that make it one of the most likely candidates for realising new and exotic phases of matter, inter alia, quantum spin liquid states: it harbors both promise and obstacles, albeit in a somewhat imbalanced way. 

In the view of all the abovementioned difficulties,
the best source for information on these low temperature phases are experiments. 
In particular since scattering neutrons off a frustrated magnet is a priori no more involved than off an unfrustrated one, and $d=3$ permits the straightforward study of bulk samples.

Up to now 
several compounds are known whose magnetic properties can be described by 
Heisenberg-like models on the pyrochlore 
lattice,\cite{daniel_kagome_2005,Rau_Gingras_review_2019,Gingras_2014}. 
While a good realisation of an $S=1$ Heisenberg magnet is now available\cite{plumb_continuum_2019}, 
 an isotropic nearest-neighbour--dominated $S=1/2$ material still remains on the wish-list. 

As experiments necessarily involve non-zero temperatures, theory is compelled to study this setting. Our study is therefore 
devoted to the $S=1/2$ pyrochlore magnet at finite temperature. We focus on the thermodynamics --  susceptibility and in particular specific heat, Fig.~\ref{fig:specheat_exp}, and we also consider the spin correlations in the form of the momentum resolved static structure factor. 

The remainder of this account is structured as follows. In Sec. \ref{sec:summary} we provide a summary of our results. Sec.~\ref{sec:model} introduces the model and observables. The bulk of the technical advances are bundled into Sec.~\ref{sec:methods}, with material on canoncial typicality, the numerical linked cluster expansion, exact diagonalisation as well as DMRG. This may be skipped on first reading (as well as by the reader not interested in the underlying methodology). Our results on thermodynamics (specific heat, susceptibility in zero and nonzero fields) as well as spin correlators are presented in Sec.~\ref{sec:results}. We close with a broader discussion and an outlook in Sec.~\ref{sec:conclusion}.

\subsection{Summary of Results}
\label{sec:summary}
\subsubsection{Methods}
Our quest to make progress has a considerable purely technical component. 
This involves efforts along two, principally computational, axes.

First, we  devise a high-order numerical linked cluster expansion (NLCE)\cite{rigol_nlce_kagome_2006,rigol_nlce_kagome_square_tri_2007,tang_nlce_2013} for the pyrochlore lattice. This approach has been used 
before, and our contribution is to push the expansion -- based on 
tetrahedral clusters well-suited to the corner-sharing tetrahedra of the pyrochlore lattice\cite{applegate_nlce_pyrochlore_exp_comp_2012,singh_nlce_tetra_pyrochlore_2012,benton_nlce_pyrochlore_2018,benton_nlce_pyrochlore_qsl_2018,jaubert_nlce_pyrochlore_exp_comp_2015,hayre_nlce_pyrochlore_2013} -- to significantly higher orders. We reach clusters of up
to eight tetrahedra, involving the full solution of clusters of up to 25 spins.
Going to such high order
allows a significantly improved exploration of the low temperature regime, in particular permitting a more extended and controlled
use of Euler transforms to extrapolate the results to low temperature. 

Indeed, high expansion orders are essential to capture a range of physical processes. Concretely, up to the sixth nearest-neighbour hop, the pyrochlore (and, indeed, the kagome lattice) is equivalent to a Husimi cactus [a Cayley tree
of tetrahedra (triangles)], and many series expansions are `trivial' to high orders, e.g.\ with degeneracy lifting
only occurring at {\it eighth} order in perturbation theory   in a high-temperature kagome\cite{harriskallinsky} 
or a  strong-coupling
pyrochlore expansion\cite{schmidt_pyro_series}. In a similar vein, the importance of resonance processes on 
more extended clusters is a recurring theme in the study of frustrated magnets. The clusters
included in our expansion host not only the simplest hexagonal loop motifs but also  loops of eight spins and longer decorated versions thereof (the longest loop consisting of eight tetrahedra cf. Fig. \ref{fig:NLCE_8tetra}), which crucially encode the three dimensional structure of the lattice. 
 
The second technical axis involves a finite-temperature DMRG analysis of the pyrochlore $S=1/2$ magnet using finite clusters. Our results demonstrate that finite-temperature DMRG is a powerful approach, feasible even in this challenging three dimensional setup down to nontrivial temperatures.
Here, we use a ``snake'' path through the lattice to map the system to one dimension with long range interactions. We take advantage of the $\mathrm{SU(2)}$ symmetry of 
the model and keep $\mathrm{SU(2)}$ block states up to $\chi=10000$ ($\sim 40000$ 
$\mathrm{U(1)}$ equivalent) and consider clusters up to 48 sites with periodic 
boundaries.

Taken together, these approaches permit us to reach converged results at temperatures down to around $T=0.25$ (where
the exchange constant of the Heisenberg model has been set to unity, Eq.~\ref{eq:xxz}) for the NLCE, and down to $T=0.6$ for DMRG.

\subsubsection{Observables}
In the zero-field specific heat we resolve a pronounced
maximum at $T = {0.57}$. Crucially, at $T=0.25$, where the specific heat has dropped well below its maximum value, the residual entropy is still around
0.33 $k_B$, i.e.  47\% of the value of a free spin of $ k_B \ln 2$. This demonstrates the persistence of the spectral weight downshift characteristic of frustrated magnets to this case. We discuss implications of this observation in detail, in particular in comparison with the kagome magnet, as well as two simple tetrahedral models, on top of the 
experimental results on two pyrochlore magnets:  the $S=1$ Heisenberg antiferromagnet 
NaCaNi$_2$F$_7$ \cite{plumb_continuum_2019}, and the classic Ising spin ice Dy$_2$Ti$_2$O$_7$ \cite{ramirez_zero-point_1999}. We find that our model at  $T=0.25$, Fig.~\ref{fig:specheat_exp},
 {\it has a higher low-$T$ entropy than all of these.}

This entropy at $T=0.25$ is in particular much greater than that proposed for singlet subspaces in resonating valence bond type effective theories. Indeed, in this regime, there is considerable admixture of triplet components in each tetrahedral wavefunction, reflecting the inability of neighbouring tetrahedra to be in singlet states simultaneously. In the magnetisation curves, this is reflected in a non-monotonic temperature dependence for fixed intermediate fields: upon cooling from the maximum magnetisation, the entropy of the magnetic excitations loses out to the singlet-dominated low-energy sector; while at high temperatures, the magnetisation assumes a conventional asymptotic $1/T$ behaviour.  The maximum disappears at zero field, where 
there is no magnetisation in the absence of time-reversal symmetry breaking; and at high fields, where a conventional
monotonic paramagnetic magnetisation curve is found. 

For the magnetic-field dependence of the specific heat, we find a continuous drift of the location of its maximum to higher temperatures; at the same time, the amplitude of the maximum changes non-monotonically, first decreasing and then increasing again.  

The spin correlators in turn, exhibit the by now familiar structure of incipient bow-ties, commonly found in various 
magnets on the pyrochlore lattice. These reflect the emergent gauge field and while they become arbitrarily 
sharp in the cases of classical 
magnets, their finite width indicates the presence of a nonzero net moment on the tetrahedra, on account of the abovementioned inability to have tetrahedra sharing a spin to be in a singlet configuration simultaneously.

\section{Model and observables}
\label{sec:model} 

We focus on the isotropic spin $S=\frac 1 2$ Heisenberg antiferromagnet

\begin{equation}
    H = \sum_{\langle i,j \rangle} \vec{S}_i \cdot \vec{S}_j  + h\sum_i S^z_i.
    \label{eq:xxz}
\end{equation}

The spins reside on the sites $i$ of the 3d pyrochlore lattice, which is a face centered cubic lattice with lattice vectors $\vec{a}_1=\frac{1}{2}(1,1,0)^T$, $\vec{a}_2=\frac{1}{2}(1,0,1)^T$, $\vec{a}_3=\frac{1}{2}(0,1,1)^T$ and a tetrahedral basis given by $\vec{b}_0=(0,0,0)^T$, $\vec{b}_1=\frac{1}{4}(1,1,0)^T$,  $\vec{b}_2=\frac{1}{4}(1,0,1)^T$,  $\vec{b}_3=\frac{1}{4}(0,1,1)^T$, such that each lattice point can be expressed by
\begin{equation}
    \vec{R}_{\alpha,n_1,n_2,n_3} = n_1 \vec{a}_1 +  n_2 \vec{a}_2 +  n_3 \vec{a}_3 + \vec{b}_\alpha,
    \label{eq:latticevectors}
\end{equation}
with integer $n_1,n_2,n_3$ and $\alpha \in \{0,1,2,3\}$.

The sum $\langle i, j\rangle$ in Eq. \eqref{eq:xxz} runs over nearest neighbor bonds of the pyrochlore lattice. In the absence of a magnetic field $h=0$, the model is SU(2) symmetric.

In this work, we focus on thermodynamic observables: the heat capacity at fixed volume $C_V$, the magnetic susceptibility $\chi$, the entropy $S$. We also consider the static spin structure factor $S(\vec{Q})$. In the following definitions, we use the canonical ensemble averages $\langle \bullet \rangle_\beta = \frac{1}{Z} \tr\left( \E^{-\beta H} \bullet \right).$ As usual, $\beta=1/T$ denotes the inverse temperature, and $Z=\tr\left(\E^{-\beta H}\right)$ is the partition function. We use natural units $k_B=1$, $\hbar=1$, $J=1$.

The  heat capacity is obtained either from the temperature derivative of the internal energy $\langle H \rangle_\beta$ or from the fluctuations of the energy:
  \begin{equation}
      C_V = \frac{\partial \langle H \rangle_\beta}{\partial T} =  \beta^2 \left(\langle H^2 \rangle_\beta - \langle H \rangle_\beta^2 \right). 
      \label{eq:spec_heat}
  \end{equation}

  Similarly, we obtain the magnetic susceptibility $\chi$, defined by the change of the magnetization in $z$ direction with respect to a change of the field $h$ in $z$ direction from the fluctuations of the magnetization:
  \begin{equation}
  	\begin{split}
        &\chi =  \frac{\partial \langle m_z \rangle_\beta}{\partial h} = \beta \left(\left\langle m_z^2 \right\rangle_\beta - \left\langle m_z \right\rangle_\beta^2 \right),\\
      &\mathrm{with }, m_z = \sum_i S_i^z
      \label{eq:chi}
     \end{split}
  \end{equation}
  In the $\mathrm{SU(2)}$-symmetric case, the susceptibility can be also 
expressed as:
\begin{equation}
 \chi=\frac{\beta}{3N}\sum_{ij}\langle \vec{S}_i \cdot \vec{S}_j \rangle.
\end{equation}

  The thermodynamic entropy $S$ can be calculated using the definition of the free energy:
  \begin{equation}
      S = \ln Z + \beta \langle H \rangle_\beta 
      \label{eq:entropy}
  \end{equation}

The static structure factor can be obtained from the Fourier transformation of 
the spin-spin correlations (the factor $4/3$ stems from the normalization $1/(S(S+1))$ for spin $S=1/2$):
\begin{equation}
    S(\vec{Q})= \frac{4}{3N} \sum_{ij} \langle \vec{S}_i\cdot 
\vec{S}_j\rangle_{\beta} \cos\left[\vec{Q}\cdot 
\left(\vec{R}_i-\vec{R}_j\right)\right],
\end{equation}
where $\vec{R}_i$ denote the real-space coordinates of sites according to Eq. \eqref{eq:latticevectors}.
  
\section{Methods}
\label{sec:methods}
This section is devoted to a detailed exposition of the methods used to obtain the results presented in the following section. It can safely be skipped at first reading, as well as by the reader primarily interested in the behaviour of the observables, rather than details on how they were obtained. 

\subsection{Canonical typicality}

Calculating thermodynamic expectation values is possible via the density matrix $\rho_\beta = \frac{1}{Z} \E^{-\beta H}$, which can be calculated from all eigenvalues and eigenvectors of the Hamiltonian $H$.
Due to the exponential scaling of the Hilbert space dimension with system size, this is impractical for systems with more than $\gtrsim 25$ spins (we discuss how to perform full diagonalization for such systems in Sec. \ref{sec:fulled}). The concept of quantum typicality \cite{popescu_entanglement_2006,goldstein_canonical_2006} permits a different approach, which has its foundation in  L\'evy's lemma. It can be summarized in the statement that for the vast majority of wavefunctions $\ket{\psi}$, the state $\ket{\beta} = \E^{-\beta/2 H}\ket{\psi}$ is typical\cite{sugiura_thermal_2012,sho_typicality_2013,luitz_ergodic_2017} for the canonical ensemble.

In practise, this means that, starting from a random wavefunction $\ket{\psi}$, one can calculate finite temperature expectation values of observables $O$ by 
\begin{equation}
    \frac{\tr(\E^{-\beta H} O)}{Z} = \frac{\bra{\beta} O \ket{\beta}}{\langle\beta|\beta\rangle} + \mathcal{O}(\E^{-N}).
\end{equation}
The statistical error of this replacement is exponentially small in system size \cite{schnack_finite-size_2020} and can be estimated (and reduced) by sampling over random (infinite temperature) initial wavefunctions $\ket{\psi}$.
The application of $\E^{-(\beta/2) H}$ corresponds to imaginary time evolution up to $\beta/2$ (the factor $1/2$ stems from a symmetric splitting of the exponential) and can be carried out efficiently using Krylov space techniques \cite{park_unitary_1986,knizhnerman_calculation_1991,jaklic_lanczos_1994,moler_nineteen_2003,luitz_ergodic_2017}, which is commonly known as the finite temperature Lanczos method \cite{jaklic_lanczos_1994}. The main advantage of this technique is that it can be carried out storing only the vectors of the Krylov space spanned by $\ket{\psi}$, while the application of the Hamiltonian to vectors during the Lanczos algorithm is either performed using a sparse matrix representation, or an on the fly generation of the corresponding matrix elements, making very large system sizes accessible which are comparable to Lanczos ground state calculations.

We note that the same techniques for reducing the Hamiltonian to its symmetry sectors discussed in Sec. \ref{sec:fulled} can be readily applied here.

\subsection{Finite temperature calculations with matrix product states}
Matrix-product-state (MPS)\cite{McCulloch_2007,schollwock_review_2011} based 
algorithms also provide a way to 
address the equilibrium thermodynamics of many-body quantum 
systems.\cite{cirac_purification_2004,white_metts_2009,Stoudenmire_2010,
white_purification_2005} One of the 
most widely used methods is the purification of the finite-temperature density 
matrix.\cite{cirac_purification_2004,white_purification_2005,bruognolo_matrix_2017} The idea behind 
this approach is that one can interpret the density 
matrix, $\rho_P$, as a partial trace of a Schmidt decomposition of a pure 
state, $|\Psi\rangle$, in an enlarged Hilbert space:
\begin{equation}
 |\Psi\rangle=\sum_{\alpha} s_{\alpha}|\alpha\rangle_P| \alpha\rangle_A 
\quad\rightarrow\quad \rho_P={\rm Tr}_A |\Psi\rangle\langle \Psi|, 
\end{equation}
where $P$ and $A$ denote the physical and auxiliary system, respectively.
The state $|\Psi\rangle$ can be easily constructed by 
simply creating a copy (auxiliary system) of the physical system and generating 
maximally entangled bonds between each physical site and its auxiliary site. 
This can be achieved by creating an entangler Hamiltonian whose ground state 
corresponds to the maximally entangled initial state. In our case this 
Hamiltonian simply reads:
\begin{equation}
    H_{\rm entangler} = \sum_{i} \vec{S}_i \cdot \vec{S}_{a(i)},
    \label{eq:xxzent}
\end{equation}
where $a(i)$ denotes the auxiliary site belonging to site $i$ and the sum is 
performed over the physical sites. One can easily see that the density matrix 
calculated from this ground state, $|\Psi_{\beta=0}\rangle$, corresponds to 
infinite-temperature.
Any 
finite-temperature density matrix can be obtained by performing an imaginary 
time evolution on the physical system and tracing out the auxiliary degrees of 
freedom: 
\begin{equation}
 |\Psi_{\beta}\rangle=e^{-\beta H/2} |\Psi_{\beta=0}\rangle 
\quad\rightarrow\quad \rho_{\beta}={\rm Tr}_A |\Psi_{\beta}\rangle\langle 
\Psi_{\beta}|.
\end{equation}
As a matter of fact, any expectation $\langle O \rangle_{\beta}$ can be 
directly evaluated from $|\Psi_{\beta}\rangle$:
\begin{equation}
 \langle O \rangle_{\beta}=\frac{\langle\Psi_{\beta}| O   
|\Psi_{\beta}\rangle}{\langle\Psi_{\beta}|\Psi_{\beta}\rangle}.
\end{equation}
This provides a great advantage of this method, since thermodynamic quantities can directly be
obtained by simulating the density matrix, rather than averaging over low entanglement pure
states \cite{white_metts_2009,Stoudenmire_2010}, and therefore results are free of statistical errors.
By design, this method is most efficient 
in one dimension. In order to use it for a three-dimensional system, one has to 
put a 'snake' path through the lattice sites to map the original problem to a 
one-dimensional equivalent one, which contains long-range couplings between the 
lattice sites. This is the main difficulty of this approach, since the MPSs 
need to encode a large amount of entanglement, 
i.e. if we think of the area law for a moment (valid only for ground states), 
the bond dimension should scale exponentially in  $\ell^2$ 
 ($\ell$ is the linear size of the 3D system) to accurately represent the 
many-body state. This obviously limits the feasible system sizes and the 
accessible temperatures.
The presence of long-range couplings in the one-dimensional topology poses 
another difficulty regarding the choice of the time evolution 
method.\cite{hubig_review_2019} The 
time-evolving block 
decimation (TEBD)\cite{vidal_2004,Daley_2004} is very effective if the couplings 
are short-ranged, otherwise one has to subsequently apply a series of swap gates to move distant 
sites next to each other so that the time evolving operator can be applied. 
This swapping procedure is extremely slow and becomes very inefficient as the 
bond 
dimension is increased. The Krylov method\cite{Garcia_Ripoll_2006} is capable of 
handling long-range 
interactions by default, but already in the early stages of the imaginary-time 
evolution the Krylov vectors become strongly entangled, making the 
calculation unfeasible. To reach physically relevant temperatures, we 
demonstrate that the time-dependent variational principle 
(TDVP)\cite{haegeman_2011,haegeman_2016} provides an 
effective way. Although it introduces another source of error by projecting the 
evolution vector onto the MPS manifold, this error is usually much smaller than 
the 
truncation error. At this point we also have to mention that it is not 
straightforward to evolve $|\Psi_{\beta=0}\rangle$ directly with 
TDVP.\cite{hubig_review_2019} Since 
the initial state is essentially a product state a naive evolution with TDVP 
would be simply wrong due to the loss of long-range interactions already in the 
first projection step. To overcome this difficulty we apply the same trick that 
has been successfully applied in real-time evolution,\cite{hubig_review_2019} 
namely, we 
generate an initial state with an artificially enlarged bond dimension.
This is achieved by finding the ground-state of the Hamiltonian:
\begin{equation}
 H_{\rm DMRG} = H_{\rm entangler} + aH(h=0)
\end{equation}
where the parameter $a$ is being varied. We start with $a=1$ and perform 20 
sweeps then we reduce it by a factor of ten each time. During the sweeps 
single-site version\cite{hubig_2015} of the density-matrix renormalization group 
(DMRG) 
algorithm\cite{white_1992,white_1993,schollwock_review_2011,hallberg_review} is 
applied with 
subspace expansion and setting the truncation error to zero. Five stages are 
performed altogether. In the last stage we set $a=0$ and perform three 
additional sweeps. We demonstrate that this procedure removes the above 
bottleneck of TDVP also for imaginary-time evolution.
In addition, to encode the large amount of entanglement, the 
compression of the many-body states must be very efficient. To this end we 
exploit the $\mathrm{SU(2)}$ symmetry of the model and keep block states up to 
$\sim 10000$ 
($\sim 40000$ U(1) equivalent) to minimize the truncation error as much as 
possible.\cite{hubig:_syten_toolk,hubig17:_symmet_protec_tensor_networ} It is 
worth mentioning that higher bond dimensions can be achieved in a ground-state 
search,\cite{qin_2019} where one can take advantage of single-site DMRG as well 
as 
parallelizing the computation in real space to reduce memory usage and 
computation time respectively. In our case, however, the two-site TDVP 
update-scheme needs to be used and the serial solution of the TDVP equations is 
crucial.
 
\subsection{Numerical linked Cluster expansion}
For studying three dimensional frustrated quantum magnets 
most controlled algorithms are restricted to a small number of spins and thus have a hard time  
capturing the three dimensional structure and its correlations. 
Using a systematic high temperature series expansion opens up the possibility to obtain 
reliable results in the thermodynamic limit. 
The numerical linked cluster expansion (NLCE) is able to determine any 
extensive property $P$ in the high temperature regime. NLCE has been applied to 
various geometries like the square lattice, kagome lattice or pyrochlore lattice 
and has provided new insights in these systems\cite{rigol_nlce_square_2007, 
rigol_nlce_kagome_square_tri_2007, tang_nlce_2013, rigol_nlce_kagome_2006, singh_nlce_tetra_pyrochlore_2012,khatami_numerical_2012} such as the 
transition of different phases in quantum 
systems\cite{hayre_nlce_pyrochlore_2013, benton_nlce_pyrochlore_qsl_2018, 
pardini_nlce_pyrochlore_qsl_2019} or a deeper understanding of real 
materials\cite{applegate_nlce_pyrochlore_exp_comp_2012, 
jaubert_nlce_pyrochlore_exp_comp_2015, benton_nlce_pyrochlore_2018}. Moreover, 
the generality allows the application of this algorithm to a variety of other
systems\cite{khatami_nlce_optical_lattic_2011, khatami_nlce_hubbard_2011, 
khatami_nlce_checkerboard_2011, khatami_nlce_clinoatacamite_2011, 
khatami_nlce_pinwheel_kagome_2011}.\par
In general, the systematic expansion can be applied to any lattice, the crucial part is the choice of building block, which builds up the infinite lattice by translational symmetries $t$. All generated configurations are extended by adding the building block in each step of the expansion. There are two main aspects that need to be considered for the choice of building block. First, the number of generated clusters scales exponentially with the order of the expansion. Second, the complexity of solving these clusters scales exponentially with systems size, which limits the maximal size of practically solvable clusters. Choosing a building block with a large number of sites induces a relatively small number of clusters which are still solvable; a building block with a small number of sites induces a very large number of solvable clusters to a degree that one may not be able to reach the largest solvable cluster size. Common choices of the building block in the square lattice is a single site or a complete square. Physically motivated, most NLCE approaches in the pyrochlore lattice use a tetrahedra expansion, based on clusters of complete tetrahedra, such that no dangling spins or triangles occur. 
In the following, we discuss the cluster expansion for multisite unit cells and compare expansions in the pyrochlore lattice based on three different building blocks, in particular the \textit{single site expansion}, the \textit{unit cell expansion} and the \textit{tetrahedra expansion}. The latter turns out to yield the most reliable results and represents the optimal expansion in that our results include full exact diagonalization of all clusters consisting of up to 8 tetrahedra. The largest included clusters thus consist of $25$ spins $\frac{1}{2}$, which host crucial loops of 6 and 8 spins in the lattice.

\subsubsection{Basic recipe}
NLCE generates all possible subclusters $c$ (subject to the choice of building blocks) which are embedded in the infinite lattice structure $\mathcal{L}$ and contribute to the thermodynamic limit $P(\mathcal{L})/N$ per site. The contribution of each cluster $c$ is given by its weight $W_P(c)$, describing the new (i.e. not included at lower order) contribution of $c$ to $P$, and its multiplicity $L(c)$, describing the number of possible embeddings of $c$ in $\mathcal{L}$.\par
The generality of this idea allows the definition of various building blocks. For the pyrochlore lattice studied here, we include a) all clusters built from single lattice sites b) all clusters built from complete (tetrahedral) unit cells and c) all clusters built from complete tetrahedra.\par
All possible configurations of clusters, subject to the choice of building blocks, are expanded further in each step of NLCE. The initial cluster given by the building block (or multiple clusters given by inequivalent building blocks) have to respect translational symmetries $t$ and cover the whole infinite structure $\mathcal{L}$ by applying these symmetries $t$.  It is important to point out the crucial difference between translational $t$ and non-translational symmetries $s$ such as rotation or reflection.\par
Each step of the expansion (at expansion order $n$) generates a set of connected clusters $\mathcal{C}_n$. This large set can be reduced to the set of clusters which are not related by lattice symmetries $\mathcal{S}_n$, and subsequently to topologically distinct clusters $\mathcal{T}_n$ of size $n$ (number of building blocks). 
The set of connected clusters $\mathcal{C}_n$ includes all possible clusters which are embedded in $\mathcal{L}$ that are not related by any translational symmetry $t$ to each other. The set of clusters not related by lattice symmetries $\mathcal{S}_n$ are given by a subset of $\mathcal{C}_n$; each cluster in $\mathcal{S}_n$ is neither related by translational $t$ nor non-translational symmetry $s$ to each other. Applying all non-translational symmetries $s$ to $\mathcal{S}_n$ generates all connected clusters $\mathcal{C}_n$. 
Moreover, the set  $\mathcal{S}_n$ can be reduced further. Even though clusters are not related by any symmetry they can exhibit the same interaction topology and hence, generate the same Hamiltonian matrix. 
We describe each cluster by its interaction graph $G$, where its nodes $i\in\mathrm{N}_G$ correspond to the spins included in the cluster and its edges $(i,j)\in\mathrm{E}_G$ correspond to nearest neighbor interaction terms of the Hamiltonian \eqref{eq:xxz}. Two clusters are topologically equivalent if there is a graph isomorphism $\pi:\mathrm{N}_{G_1}\rightarrow\mathrm{N}_{G_2}$ (bijective) mapping $G_1$ on $G_2$ while preserving its structure; that means if $(i_1, j_1)\in\mathrm{E}_{G_1}$ is an edge of $G_1$ then $(\pi(i_1), \pi(j_1))\in\mathrm{E}_{G_2}$ needs to be an edge of $G_2$. Hence, the set of topologically distinct clusters $\mathcal{T}_n$ is a subset of $\mathcal{S}_n$. Using building blocks including more than one lattice site (like the tetrahedron) requires to check the topological structure on the full connectivity graph including all sites.\par
The multiplicity $L(c)$ assigned to each cluster $c$ describes the number of possible embeddings in the infinite structure $\mathcal{L}$. All possible clusters (subject to the choice of building blocks) of size $n$ are included in $\mathcal{C}_n$; hence, the multiplicity of each cluster is one. Non-translational symmetries $s$ reduce the set of connected cluster to $\mathcal{S}_n$ and summarize multiple clusters in $\mathcal{C}_n$ to one representative cluster $c\in \mathcal{S}_n$. The multiplicity of each cluster in $\mathcal{S}_n$ is given by the number of non-translational symmetries $s$ that transform the cluster to another cluster that is not related to the first one by any translation $t$. Again, multiple clusters in $\mathcal{S}_n$ are summarized to one representative cluster $c\in \mathcal{T}_n$, its multiplicity given by the number of topologically equivalent clusters. Hence, the multiplicity of topologically invariant clusters is simply the sum of the multiplicities of all topologically equivalent clusters in $\mathcal{S}_n$. Summing all multiplicities of clusters in $\mathcal{S}_n$ or $\mathcal{T}_n$ equals the number of connected clusters:
\begin{equation}
    \sum_{c\in\mathcal{S}_n} L_\text{sym}(c) = \sum_{c\in\mathcal{T}_n} L_\text{top}(c) = \vert\mathcal{C}_n\vert.
    \label{eq:nlce_00}
\end{equation}
For clarity, we will drop the index $\text{top}$ in what follows, i.e. $L(c) = L_\text{top}$ etc.
The basic recipe to expand the clusters by one building block ($n\,\rightarrow\,n+1$) is equivalent for all geometries and building blocks: 
\begin{enumerate}
	\item[\textbf{i)}] Starting from all clusters not related by lattice symmetries of size $n$ in $\mathcal{S}_n$, we generate new clusters by adding a building block to every free nearest neighbor. Again, we only consider clusters that are distinguishable by translational symmetries $t$.
  	\item[\textbf{ii)}] Non-translational symmetries $s$ are used to reduce the set of all newly generated expansions to create the set $\mathcal{S}_{n+1}$. Applying all non-translational symmetries $s$ to $S_{n+1}$ generates the full set of connected clusters $\mathcal{C}_{n+1}$ with respect to translational equivalence.
  	\item[\textbf{iii)}] Clusters in $\mathcal{S}_{n+1}$ are further reduced to obtain all topologically distinct clusters in $\mathcal{T}_{n+1}$.
\end{enumerate}
Each cluster in $\mathcal{T}_n$ contributes to the expansion according to its multiplicity $L(c)$ and weight $W_P(c)$. The $n$th order of NLCE is given by:
\begin{equation}
	\left. P(\mathcal{L})/N \right|_n = \sum_{m=0}^n\sum_{c\in \mathcal{T}_m} L(c)W_P(c)\label{eq:nlce_01}
\end{equation}
The weight assigned to a cluster $c$ is defined with respect to all smaller subclusters $s\subset c$ (subject to the choice of building blocks) which can be embedded in $c$; hence, it extracts contributions of $c$ to $P$ which are not covered by smaller clusters: 
\begin{align}
	W_p(c) := P(c) - \sum_{s\subset c} W_P(s)\label{eq:nlce_02}
\end{align}
In practice, the thermodynamic observable $P$ needs therefore to be calculated for all topologically invariant clusters. The extensive property of $P$ induces a zero weight for disconnected clusters\cite{tang_nlce_2013} which do not have to be considered in \eqref{eq:nlce_02}. Expanding clusters only by nearest neighbors guarantees connected clusters in $\mathcal{C}_n$, $\mathcal{S}_n$ and $\mathcal{T}_n$. The recursive definition of the weight ensures the convergence towards the infinite structure $\mathcal{L}$ which is the thermodynamic limit.\par
It is not necessary to use all (or any) non-translational symmetries in the NLCE. A lower number of non-translational symmetries increases the number of clusters $\vert\mathcal{S}_n\vert$ such that each cluster has a lower multiplicity. In fact, it is possible to consider the identity $\mathrm{id}$ as the only non-translational symmetry, then $\mathcal{C}_n = \mathcal{S}_n$. However, checking the topological structure of these clusters generates the same set of topologically distinct clusters $\mathcal{T}_n$. The computational effort can increase drastically with a low number of symmetries, since the number of clusters grows exponentially.
\subsubsection{Pyrochlore lattice and building blocks}
The underlying Bravais lattice is given by a fcc-structure with a tetrahedral unit cell (four sites). Hence, in order to respect the translations, all expansions we use focus on the tetrahedral unit cell and converge to the thermodynamic limit per unit cell and not per site, thus accounting for an extra factor of four. \par
Its symmetries are described by the space group $Fd\bar{3}m$ (227); it contains 192 symmetry operations. However, only 12 symmetries are purely non-translational and used in the NLCE, see table \ref{tab:nlce_00}.\par
\begin{table}
	\centering
		\begin{tabular}{ l l }
			\hline\hline
			$s_1$: $\mathrm{id}$ & $\qquad s_7$: ${2\, (0,\bar{y},y)}$\\
			$s_2$: $3^+\, (x,x,x)$ & $\qquad s_8$: ${2\, (\bar{x},0,x)}$\\
			$s_3$: $3^-\, (x,x,x)$ & $\qquad s_9$: ${2\, (x,\bar{x},0)}$\\
			$s_4$: $\mathrm{m}\,(x,y,y)$ & $\qquad s_{10}$: ${\bar{1}\, (0,0,0)}$\\
			$s_5$: $\mathrm{m}\,(x,y,x)$ & $\qquad s_{11}$: ${\bar{3}^+\, (x,x,x;0,0,0)}$\\
			$s_6$: $\mathrm{m}\,(x,x,z)$ & $\qquad s_{12}$: ${\bar{3}^-\, (x,x,x;0,0,0)}$\\
			\hline\hline
		\end{tabular}
		\caption{Non-translational symmetries in the pyrochlore lattice.}
	\label{tab:nlce_00}
\end{table}
$s_2$, $s_3$ and $s_7$, $s_8$, $s_9$ describe three and two folded rotations, respectively. Reflections are given by $s_4$, $s_5$, $s_6$ and $s_{10}$ represents the inversion. $s_{11}$ and $s_{12}$ combine a threefold rotation with the inversion.\par
As discussed earlier, the choice of building block is crucial. In principle, various geometries (such as dimers, hexagons or multiple tetrahedra) can be used as building blocks as long as they respect the translations $t$ and cover the whole lattice. Each expansion we use is embedded in the fcc structure of the pyrochlore with either equivalent (unit cell expansion) or multiple inequivalent (single site and tetrahedra expansion) building blocks.\par
An efficient implementation of translational symmetries is essential due to the exponentially increasing complexity. Labeling lattice sites along the translational axes automatically describes the translational symmetries $t$ by a simple index shift.\par
\textbf{a)} \textit{single site expansion}:\par
As pointed out earlier, the single site expansion generates a large number of clusters of relatively small size. The advantage of this approach is its complete generality. In contrast to the single site expansion in Bravais lattices (e.g. square or triangular lattice\cite{rigol_nlce_square_2007, rigol_nlce_kagome_square_tri_2007, tang_nlce_2013}), the unit cell in the pyrochlore lattice consists of four sites, which need to be treated inequivalently. Hence, the starting point of the single site expansion are four sites arranged in the unit cell/tetrahedron, which covers the whole lattice by translations. Applying translation symmetries to these four sites generates the full pyrochlore lattice. Here, all symmetries in table \ref{tab:nlce_00} can be applied to find clusters which are related by lattice symmetries to reduce the complexity. \par
\textbf{b)} \textit{unit cell expansion}:\par The unit cell expansion is related to the single site expansion in the 
fcc-lattice, substituting each site in the obtained clusters by the tetrahedral 
unit cell. However, working within the pyrochlore lattice, the symmetries in 
table \ref{tab:nlce_00} have to be examined more carefully: We require the 
symmetries to preserve the unit cell structure such that only entire unit cells 
are mapped to each other. Only the symmetries $s_1$ to $s_6$ in table 
\ref{tab:nlce_00}, which are a subset of the fcc-lattice symmetries, are unit 
cell conformal. As mentioned before, working with a lower number of symmetries 
produces the same results. Since the building block includes more than one site, 
the topological structure has to be compared on the level of the full 
connectivity graph including all lattice sites. The advantage of this approach 
is the consideration of larger clusters due to a much slower growth of the 
number of clusters with the number of unit cells. However, the connection 
between the unit cells are dangling bond that do not reflect the geometrical 
frustration. In the presence of magnetic fields, the Hamiltonian has bond and site
terms. Therefore, we include a single site (yielding the simplest contribution of the site terms) 
as the $0$th order in the expansion ($m=0$) in \eqref{eq:nlce_01}, which is embedded four times in the unit cell. \par 
\textbf{c)} \textit{tetrahedra expansion}:\par
The central motif of the pyrochlore lattice is the tetrahedron. An examination of the lattice shows that there are two types of tetrahedra: up pointing tetrahedra (these are the unit cells) and down pointing tetrahedra, which correspond to the interaction of each spin in the unit cell to three neighboring spins in \emph{different} unit cells. Both the single site and unit cell expansion do not respect this structure, leading to dangling bonds in the case of most clusters in the single site expansions and incomplete down pointing tetrahedra in the case of the unit cell expansion.\par
For this reason, we use an expansion including all clusters with \emph{complete tetrahedra}\cite{singh_nlce_tetra_pyrochlore_2012, benton_nlce_pyrochlore_qsl_2018, pardini_nlce_pyrochlore_qsl_2019, applegate_nlce_pyrochlore_exp_comp_2012, jaubert_nlce_pyrochlore_exp_comp_2015, benton_nlce_pyrochlore_2018}. The systematic expansion focuses on an hourglass structure composed by two inequivalent building blocks of thetrahedra (up/down pointing) which are placed in the underlying fcc-lattice and expands these as described before. Our comparison of results for the heat capacity and the magnetic susceptibility demonstrates that this intuition is correct and that this expansion is indeed superior (cf. comparison in Appendix \ref{sec:nlcecomp}).\par
It is important to note that the physical size (number of spins) of each cluster is not uniquely related to the order of the expansion, due to an overlap of up- and down-facing tetrahedra. This means that at an expansion order (given by the total number of tetrahedra) $n_{\text{tetra}}$ clusters of different sizes are included.
Similarly to the unit cell expansion, this expansion leads to a relatively small number of large clusters. Again, we need to consider the $0$th order contribution of a single site. We did not apply any symmetry due to the small number of edges of each tetrahedron and the low order of expansion and rely on the automatic identification of topologically equivalent clusters by directly comparing their interaction graphs.\\ \par
In appendix \ref{app:nlce} we provide a comparison of the number of clusters generated at each order in the three expansions discussed here. We list the number of connected $\vert\mathcal{C}_n\vert$, not related by lattice symmetries $\vert\mathcal{S}_n\vert$ (if present) and topologically distinct clusters $\vert\mathcal{T}_n\vert$ in tables \ref{tab:app_nlce_00}, \ref{tab:app_nlce_01} and \ref{tab:app_nlce_02}. The visualization of these results is shown in figure \ref{fig:nlce_00}, clearly showing that the number of clusters with at most $n$ sites is smallest in the tetrahedra expansion, leading to a tractable number at the edge of feasibility of full diagonalization (the maximal symmetry block dimension of the Hamiltonian is 228592) for clusters with $25$ sites (full Hiblert space $3.355\cdot 10^7$)). Additionally, all topologically distinct clusters of the tetrahedra expansion can be found in the appendix \ref{app:nlce}.

\subsubsection{Resummation algorithms}
\label{sec:resum}

Correlations are increasingly long ranged as temperature is lowered. 
Hence, contributions of larger cluster have to be taken into account to converge 
to the thermodynamic limit. Accessing these orders is limited due to the 
exponentially increasing complexity regarding the number of clusters or Hilbert 
space dimension.
\par One effective tool to obtain reliable data for lower temperature are 
resummation algorithms like Euler's transformation \cite{numerical_recipes} which can accelerate the convergence of NLCE. 
Detailed descriptions and examples can be found in 
Refs.~[\onlinecite{tang_nlce_2013,rigol_nlce_kagome_2006,rigol_nlce_square_2007,
khatami_nlce_optical_lattic_2011, khatami_nlce_hubbard_2011, 
khatami_nlce_checkerboard_2011, khatami_nlce_clinoatacamite_2011, 
khatami_nlce_pinwheel_kagome_2011}]. 

Resummation algorithms rely on a systematic usage of lower orders of the series and are most effective if many terms are included. They are guaranteed to converge to the limiting value of the underlying series, or not at all. In this work, we use the Euler transform of our NLCE data for the expansion up to $n=8$ tetrahedra and compare the highest order Euler transform of the Euler transform up to expansion orders $n=7$ (and also $n=6$), to ensure convergence of our results.

Euler's transformation is particularly useful for alternating series, which are transformed according to\cite{numerical_recipes}
\begin{equation}
    \sum_{s=0}^{\infty} (-1)^s u_s = \sum_{s=0}^{\infty} \frac{(-1)^s}{2^{s+1}} \left[ \Delta^s u_0 \right],
    \label{eq:alternating}
\end{equation}
where $\Delta^s$ is the $s$ fold application of the forward difference operator $\Delta$, defined by $\Delta u_n := u_{n+1} - u_n$. Euler's method can be derived by repeated application of summation by parts\cite{Hamming1987}.

In the present work, we find that the Euler transformation of our NLCE series up to $n=8$ tetrahedra (i.e. containing 8 terms) yields a \emph{significant improvement of convergence} at low temperatures. We always compare the Euler transformation of the first $n=7$ and $n=8$ terms of the series to ensure that the results are indeed converged, yielding reliable results down to $T\approx 0.25$ for thermodynamic properties of the $S=1/2$ pyrochlore antiferromagnet as shown in detail in what follows.

 \subsection{Exact diagonalization of clusters}
\label{sec:fulled}

\subsubsection{Cluster symmetries from graph automorphisms}
The numerical linked cluster expansion (NLCE) expresses thermodynamic 
observables  as a series expansion in terms of the exact solution of a large 
number of finite size clusters. Since the number of clusters grows factorially 
with the number of constituents, it is useful to use an automatic strategy to 
identify cluster symmetries, which are used for block diagonalizing the 
Hamiltonian. Here, we provide a practical description of the method with only 
minimal reference to graph and group theory to make it accessible. A similar 
pedagogical description for the exploitation of translational symmetry can be 
found in Ref.~[\onlinecite{sandvik_computational_2010}]. In all of the 
following discussion, we use the computational $S^z$ basis, in which each 
basis state $\ket{\sigma_1,\sigma_2,\dots}$ is an eigenstate of all local 
$S_i^z$ operators and labelled by their eigenvalues $\sigma_i$.

As described in the NLCE part, we identify a finite cluster with its interaction graph $G$, where its nodes $i\in\mathrm{N}_G$ correspond to the spins and its edges $(i,j)\in\mathrm{E}_G$ correspond to nearest neighbor interaction terms of the Hamiltonian \eqref{eq:xxz}. Hence, the Hamiltonian is defined by:

\begin{equation}
    H = \sum_{(i,j) \in \mathrm{E}_G} \vec{S}_i \cdot \vec{S}_j + h \sum_{i\in \mathrm{N}_G}  S^z_i.
\end{equation}

The sum $(i,j)\in \mathrm{E}_G$ runs over all edges $(i,j)$ of the graph $G$, the sum $i\in \mathrm{N}_G$ runs over all sites. 
We notice that any \emph{automorphism} of the graph $G$ leaves the Hamiltonian invariant, since an automorphism is a \emph{permutation $\pi$ of nodes} which maps the graph onto itself, such that for any edge $(i,j) \in \mathrm{E}_G$, the mapped edge has to be in $G$ as well: $(\pi(i), \pi(j))\in \mathrm{E}_G$.

This means that graph automorphisms are \emph{symmetries} of the Hamiltonian and commute with it, which implies that we can simultaneously diagonalize the automorphism and the Hamiltonian. 

The matrix representation $\mat{A}$ of a graph automorphism $A$ (which necessarily is a permutation $\pi_A$ of graph nodes) can be obtained by noticing that any basis state is transformed as
\begin{equation}
    A \ket{\sigma_1,\sigma_2,\dots,\sigma_N} = \ket{\sigma_{\pi_A(1)}, \sigma_{\pi_A(2)}, \dots, \sigma_{\pi_A(N)}}.
\end{equation}
$\mat{A}$ is a permutation matrix on the set of basis states and $\mat{A}^T\mat{A}=\eye$, i.e. $\mat{A}$ is orthogonal with eigenvalues on the complex unit circle. Since $\mat{A}$ is a permutation matrix, it is \emph{idempotent with a certain order $N_A<\vert N_G\vert$}: $\mat{A}^{N_A}=\eye$. Therefore, its eigenvalues are given by the $N_A$-th roots of unity: $\E^{\I 2\pi n_A/N_A}$, with $n_A\in\{0,\dots ,N_A-1\}$.

This means, we can block diagonalize $H$ into $n_A$ blocks, labelled by the eigenvalues $\E^{\I 2\pi n_A/N_A}$ of $\mat{A}$. We denote the order $o(\pi)$ of an automorphism $\pi$ by the minimal number $n\in\mathbb{N}$ such that $\pi^n = \eye$.

\subsubsection{Identification of largest commuting automorphism subgroup}

Typical interaction graphs $G$ have a large number of (independent) graph automorphisms. 
However, typically \emph{not all of them commute}! Since only commuting graph 
automorphisms can be used together to further reduce the block size of the 
Hamiltonian, we want to find the largest abelian (commuting) subgroup of the 
complete automorphism group of $G$.\par
In order to do so, we start by generating the automorphism group of the graph. In the next step, we check for each pair of automorphisms if they commute. This can be interpreted by a new graph $C$, in which each automorphism of $G$ corresponds to a node and two nodes are connected if and only if the corresponding automorphisms commute. What we are looking for is a subgroup $\mathcal{U} \subset \mathrm{N}_C$ in which each node in $\mathcal{U}$ is connected to all other nodes of the subgroup. Hence, each element in $\mathcal{U}$ commutes with each other. This is called a \emph{clique} in graph theory. Finding the largest abelian subgroup of the automorphism group is therefore identical to finding the largest \emph{clique} in $C$. In general, the largest \emph{clique} is not uniquely determined.\par
After determining the largest \emph{clique} $\mathcal{U}$ of automorphisms, we need to identify a \textit{minimal} set $\mathcal{H}$ of independent generators of the abelian subgroup. Each element of the \emph{clique} $\mathcal{U}$ has to be generated \textit{uniquely} by elements of $\mathcal{H}$. Assume, $\mathcal{H}$ includes $m:=\vert \mathcal{H}\vert$ elements $h\in\mathcal{H}$ of order $o(h)$, then there is exactly one tuple $(n_{h_1}, \dots, n_{h_m})$ of integers for each element $u$ of the subgroup, which generates $u$ from the corresponding integer powers of the generators:
\begin{align}
	&\forall u\in\mathcal{U}: \,\exists!\,(n_{h_1}, \dots, n_{h_m})\in\mathbb{N}^{m}\text{ with } 0\leq n_{h_i} < o(h_i),\nonumber\\
    &\text{such that } u = \prod_{h\in\mathcal{H}} h^{n_h}.\label{eq:multiindex}
\end{align}
The product in equation \eqref{eq:multiindex} refers to the composition of permutations. The ordering is arbitrary since all generators commute. Each element in $\mathcal{U}$ is represented uniquely by a multiindex $(n_{h_1}, ..., n_{h_m})\in\mathbb{N}^{m}$ where each number is smaller than the corresponding order of the generator. This multiindex determines a phase in the symmetrized basis (cf. Sec. \ref{sec:symbasis}). 
The bijective mapping from $\mathcal{U}$ to $\mathbb{N}^{m}$, respecting the order $o(h_i)$ of the generators $h_i$, induces the following relation between the cardinality of the abelian subgroup $|\mathcal{U}|$ and the orders of its generators:
\begin{equation}
    \vert \mathcal{U} \vert = \prod_{h\in\mathcal{H}}o(h). \label{eq:number_of_autos}
\end{equation}
In practice, we create the minimal set of generators by starting with the elements exhibiting the highest order in the commuting subgroup $\mathcal{U}$. First, one element with the highest order will be added in $\mathcal{H}$. A new element $g$ is added to $\mathcal{H}$ if it does not violate the bijective mapping described through \eqref{eq:multiindex}. That is, all possible compositions  $g^n\circ h$, for $n = 0,..., o(g) -1$ and $h\in\mathcal{H}$, generate uniquely a new element in the subgroup $\mathcal{U}$. 
The generating set $\mathcal{H}$ is not uniquely determined. 
\begin{figure}[h]
    \centering
    \includegraphics{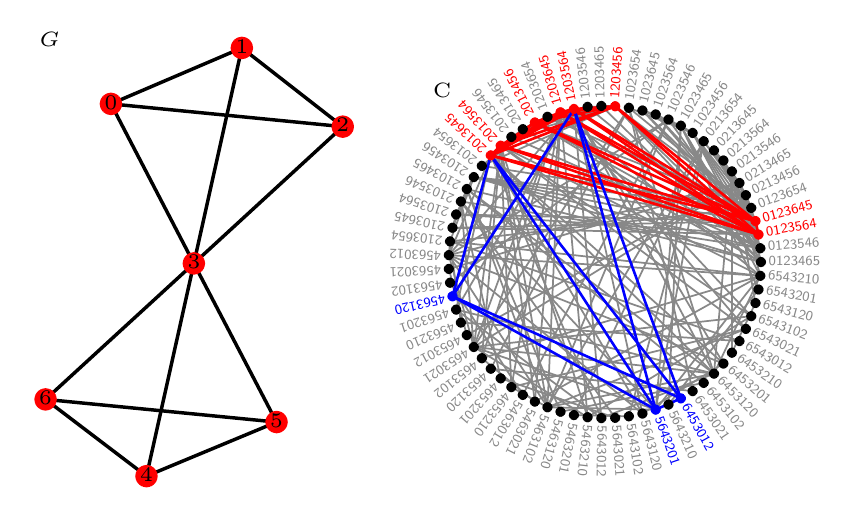}
    \caption{Left: example interaction graph $G$ of an hourglass composed by two corner sharing tetrahedra. Right: Resulting automorphism graph $C$, where each node represents one automorphism, and commuting automorphisms are connected by an edge. The red colored nodes represent the largest \emph{clique} in the graph, in which all automorphisms commute (red edges). The identity $(0,1,2,3,4,5,6)$ commutes with all automorphisms and has therefore been omitted from the graph $C$. It is part of the largest clique. The largest clique consists of the following 8 nontrivial permutations: (2, 0, 1, 3, 5, 6, 4), (1, 2, 0, 3, 6, 4, 5), (2, 0, 1, 3, 6, 4, 5), (0, 1, 2, 3, 6, 4, 5), (1, 2, 0, 3, 4, 5, 6), (2, 0, 1, 3, 4, 5, 6), (1, 2, 0, 3, 5, 6, 4), (0, 1, 2, 3, 5, 6, 4), which are marked in red, the blue edges indicate a smaller clique with 5 nontrivial permutations.
}
    \label{fig:clique_graph}
\end{figure}

In Fig. \ref{fig:clique_graph}, we show an illustration for an example interaction graph (hourglass composed of two corner sharing tetrahedra, corresponding to the $n=2$ cluster in NLCE) along with its 71 nontrivial automorphisms and their commutation relations. The largest clique for this example has 8 nontrivial automorphisms with two generators $A$ and $B$ [e.g. given by the node permutations $\pi_A=(2, 0, 1, 3, 5, 6, 4)$, $\pi_B=(2, 0, 1, 3, 6, 4, 5)$], generating independent $C_3$ rotations of the upper and lower tetrahedron. 
In the present work, we rely on established algorithms for the identification of graph iso- and automorphisms bundled in the package nauty\cite{mckay_nauty_2014} and used the clique maximization algorithm described in Ref. \cite{konc_clique_2017}.

\subsubsection{Symmetrized basis}
\label{sec:symbasis}

Once we have obtained a set of independent generators $\mathcal{H}$ and the multiindices referring to the largest abelian subgroup $\mathcal{U}$ of the graph automorphism group, we can proceed with the block diagonalization of the Hamiltonian. A subgroup of size $\vert\mathcal{U}\vert$ induces the same number of blocks, each is uniquely identified by $m$ (number of minimal generators) quantum numbers given by multiindices described in \eqref{eq:multiindex}. Each index refers to the phase of the eigenvalue of the corresponding generator; the commutation relations of the generators allows the simultaneous diagonalization of all generators and the Hamiltonian. 
Each computational basis state $\ket{\vec{\sigma}}=\ket{\sigma_1,\dots\sigma_N}$ has to be replaced by a symmetrized state induced by the quantum numbers $\varepsilon = (\varepsilon_1,...,\varepsilon_{m})\in\mathbb{N}^{m}$ which is given by
\begin{align}
	\ket{\sigma_1, \dots, \sigma_N&; \varepsilon} =\nonumber\\
	&\frac{1}{\sqrt{\mathcal{N}_{\vec{\sigma}}^\varepsilon}}\sum_{u\in\mathcal{U}} \prod_{i=0}^{m}  \E^{\I2\pi n_i^u\varepsilon_i/o(h_i)} \,u\,\ket{\sigma_1,\dots,\sigma_N},\label{eq:sym_state}
\end{align}
where $\mathcal{N}_{\vec{\sigma}}^\varepsilon$ is the normalization constant of the state, $n^u=(n_1^u,...,n_m^u)\in\mathbb{N}^m$ is the multiindex referring to $u\in\mathcal{U}$ defined by \eqref{eq:multiindex}, and $o(h_i)$ is the order of the generator $h_i\in \mathcal{H}$. The eigenvalue of each generator $h_i\in\mathcal{H}$ is obtained (using the properties of the generators and the complex roots of unity) from
\begin{equation}
    h_i\ket{\sigma_1, \dots, \sigma_N; \varepsilon} = \E^{\I2\pi \varepsilon_i/o(h_i)}\ket{\sigma_1, \dots, \sigma_N; \varepsilon}. 
\end{equation}

It is important to note that multiple unsymmetric basis states $\ket{\vec{\sigma}}$ typically generate the \emph{same} (apart from a phase) symmetric state. This is in fact true for any basis state which is in the set 
\begin{equation}
    F_{\vec{\sigma}} =   \left\{ u \ket{\vec{\sigma}} \, \Big| \, u\in\mathcal{U} \right\},
\end{equation}
which we call the ``family of the state $\ket{\vec{\sigma}}$'' generated by the commuting subgroup $\mathcal{U}$ of the graph automorphism group. Therefore, each symmetric basis state has to be added \emph{only once} to the symmetric basis. This is typically ensured by using a \emph{parent state} of the family, for example the state $\ket{\vec{\sigma}}$ with the lowest binary representation; this \emph{parent state} is denoted by $p(F_{\vec{\sigma}})$.
 
Crucially, some unsymmetric basis states $\ket{\vec{\sigma}}$ \emph{do not generate any symmetric state} in a given symmetry sector. This happens if the basis state is \emph{incompatible with the symmetry sector} and the sum of phase factors cancels, leading an unnormalizable state. 
An example is the state $\ket{\up\up\up\dots\up}$. For any graph and any automorphism $u$, it is mapped to itself. Therefore, its family is $F_{\up\up\dots\up} = \{\ket{\up\up\up\dots\up}\}$. In the symmetry sector $0^m:=(0,\dots,0)$, we obtain $\ket{\up\up\dots\up; 0^m} = \ket{\up\up\dots\up}$. In all other sectors, however, we get $\ket{\up\up\dots\up; n\neq 0^m} = 0$, i.e. this state does not appear in other sectors. As in the general example, this mechanism leads to an imbalance of the size of sectors, the sector $0^m$ always being the largest.

It is crucial to ensure the correct normalization of the symmetric basis states $\ket{\vec{\sigma}}$ and $\ket{\vec{\sigma}'}$ introduced in Eq. \eqref{eq:sym_state}. We require:

\begin{equation}
    \braket{\vec{\sigma};n}{\vec{\sigma}';n'}=  \delta_{nn'} \delta_{p(F_{\vec{\sigma}}),p(F_{\vec{\sigma}'})},
\end{equation}
i.e. states are orthogonal if they correspond to different parents (and therefore families), or if they correspond to different symmetry sectors $n$.

We note that in addition to the graph automorphisms, the Heisenberg model we study has additional spin symmetries. Here, we exploit the conservation of the total $z$ component of the spin, because $\left[\sum_i S_i^z , H\right]=0$. Since all computational basis states we use are already eigenstates of the total $z$ component $\sum_i S_i^z$, and since $\sum_i S_i^z$ also commutes with the graph automorphisms, this symmetry is trivial to exploit: A simple reordering of basis states by their $z$ magnetization makes the Hamiltonian block diagonal. Additionally, we exploit the spin inversion symmetry, $\left[Q, H\right]=0$ with $Q:=\prod_i S_i^x$ in the sector $m_z = 0$. The spin inversion is also used to deduce the results for the $-S_z$ sector from an already solved $S_z$ sector.

\subsubsection{Hamiltonian submatrix in symmetry sectors}

Before we can fully diagonalize the Hamiltonian, we need to represent each block of $H$ (labelled by the quantum numbers $\varepsilon$) in the symmetric basis. For simplicity, we only focus on the graph automorphisms and ignore symmetries defined by $S_z$ and $Q$ (which are a trivial extension). Hence, we need to construct the matrix elements  $\bra{\vec{\sigma};\varepsilon} H \ket{\vec{\sigma}';\varepsilon}$; note that by construction the inter-block matrix elements $\varepsilon\neq \varepsilon'$ are zero.\par
Let us apply the Hamiltonian to a symmetrized basis state, exploiting the fact that $H$ commutes with $u\in\mathcal{U}$:
\begin{equation}
    H \ket{\vec{\sigma};\varepsilon} = \frac{1}{\sqrt{\mathcal{N}_{\vec{\sigma}}^{\varepsilon}}}\sum_{u\in\mathcal{U}} \prod_{i=0}^{m}  \E^{\I2\pi n_i^u\varepsilon_i/o(h_i)} \,u\,H\,\ket{\vec{\sigma}}.
    \label{eq:Hpsi}
\end{equation}
The Hamiltonian is expressed as a sum of non-branching terms: $H = \sum_b h_b$; note that the permutations $u\in\mathcal{U}$ commute with each single term $h_b$. The operators $h_b$ can be divided into diagonal operators (which do not change the state $\ket{\vec{\sigma}}$) and off-diagonal operators, which yield a different state $\ket{\vec{\sigma}_b}$ together with the corresponding matrix element by $h_b\ket{\vec{\sigma}} =   h_b^{\vec{\sigma}\vec{\sigma}_b} \ket{\vec{\sigma}_b}$. $\ket{\vec{\sigma}_b}$ is in general not a parent state. It is however related to its parent state by a symmetry operation $\ket{\vec{p}_b}=u_0\ket{\vec{\sigma}_b} = p(F_{\vec{\sigma}_b})$. Note that $u_0$ is not determined uniquely; multiple permutations can fulfill this mapping. Also, the new state is not necessarily a valid state in the symmetry sector $\varepsilon$, in which case it will be canceled by later terms. The parent state is assigned to an index of the basis; the referring matrix element can be calculated as follows:
\begin{align}
	&\bra{\vec{p}_b;\varepsilon}h_b\ket{\vec{\sigma};\varepsilon} \nonumber\\
	&= \bra{\vec{p}_b;\varepsilon}\frac{1}{\sqrt{\mathcal{N}_{\vec{\sigma}}^{\varepsilon}}}\sum_{u\in\mathcal{U}} \prod_{i=0}^{m}  \E^{\I2\pi n_i^u\varepsilon_i/o(h_i)} \,u\,h_b\,\ket{\vec{\sigma}}\nonumber\\
    &= \frac{1}{\sqrt{\mathcal{N}_{\vec{\sigma}}^{\varepsilon}\mathcal{N}_{\vec{p}_b}^{\varepsilon}}}\sum_{u,u'\in\mathcal{U}} \prod_{i=0}^{m}  \E^{\I2\pi (n_i^u - n_i^{u'})\varepsilon_i/o(h_i)}\bra{\vec{p}_b}u' u\ket{\vec{\sigma_b}}    h_b^{\vec{\sigma}\vec{\sigma}_b}   \label{eq:ham_matels}
\end{align}
Terms in Eq. \eqref{eq:ham_matels} only have a non-zero contribution if and only if $\ket{\vec{\sigma}_b}$ is mapped by $u' u$ to its parent state $\ket{\vec{p}_b}$.
The matrix elements of the Hamiltonian in the symmetrized basis are therefore given by the unsymmetrized matrix elements $h_b^{\vec{\sigma}\vec{\sigma}_b}$ multiplied by symmetry sector dependent phase factors and normalization constants. We note in passing that the sums over phase factors can yield the normalization constant $\mathcal{N}_{\vec{p}_b}^\varepsilon$ \cite{sandvik_computational_2010}.
 \section{Results}
\label{sec:results}

Using the combination of the methods described in Sec.~\ref{sec:methods}, we address thermodynamic properties of the pyrochlore quantum Heisenberg antiferromagnet. 
We start by considering the SU(2) symmetric case without an applied magnetic field and present our results for the heat capacity (\ref{ssec:Cvzero}), the magnetic susceptibility (\ref{ssec:chizero}), and the thermodynamic entropy. We compare the results for different orders in the numerical linked cluster expansion (NLCE) and their Euler transform to show that the high temperature regime down to $T\approx 0.25$ is converged to the thermodynamic limit. We furthermore compare the NLCE results to the solution of finite size clusters obtained using canonical typicality and finite temperature DMRG.

We also present results obtained from DMRG for the static spin structure factor at finite temperature (\ref{ssec:sqzero}), as well as NLCE results for the  heat capacity and the magnetization at finite applied magnetic field $h$ (\ref{ssec:Cvfinite}).

\begin{figure}[h]
    \centering
    \includegraphics[width=\columnwidth]{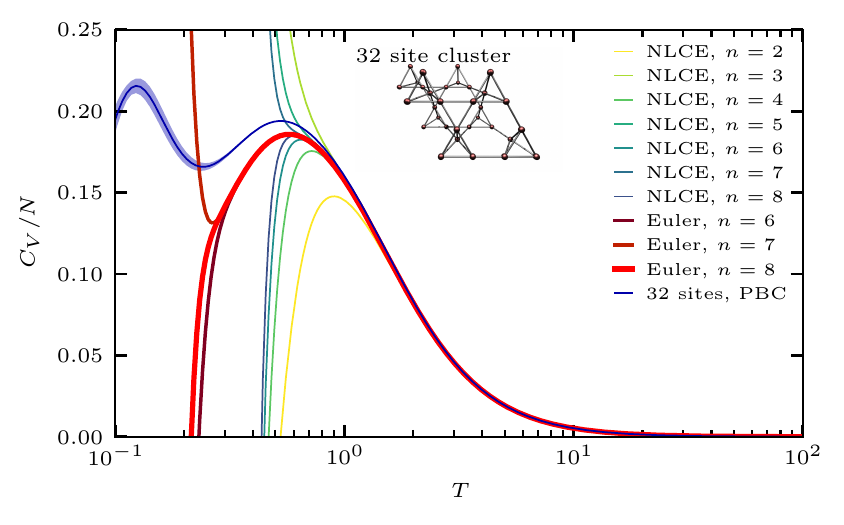}\\
        \includegraphics[width=\columnwidth]{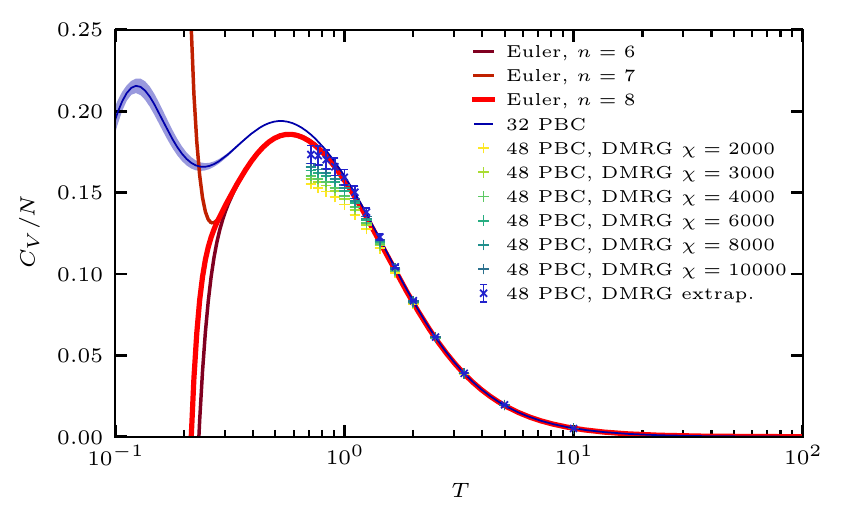}
    \caption{Comparison, and convergence, of heat capacity of the spin $1/2$ pyrochlore Heisenberg model  determined via different methods. Top: Numerical linked cluster expansion (NLCE) with clusters of $n=2\ldots 8$ complete tetrahedra, as well as their Euler transform for NLCE orders $n=6,7$, and $n=8$ (red curves). These results appear to be converged down to temperatures of about $T=0.25$. The blue curve shows the result for 32 site (8 unit cells) cluster with periodic boundaries obtained from canonical typicality. The error bars (shaded blue area) reflect variations in results obtained  from sampling over different random vectors. Bottom: DMRG results obtained from the numerical (spline) derivative of the energy $\langle H \rangle_\beta$ for different bond dimensions $\chi$ obtained by finite temperature DMRG (purification) and the blue crosses represent their extrapolation to $\chi\to\infty$ using a quadratic polynomial in $1/\chi$. The errorbars indicate the distance of the extrapolated from the $\chi=10000$ results. Typicality and Euler transform of NLCE results as in top panel.
    }
    \label{fig:spec_heat_synopsis}
        \label{fig:spec_heat_dmrg}
\end{figure}

Fig. \ref{fig:spec_heat_synopsis} shows by example of the specific heat how the converged results were obtained. The blue curve shows the  heat capacity for a finite size cluster with $N=32$ sites (8 unit cells, inset) in comparison with the results of the numerical linked cluster expansion for different orders $n$ (top panel), indicating the number of complete tetrahedra in the clusters included in the expansion. To accelerate the convergence of the NLCE series at lower temperatures, we apply the Euler transformation up to order $n$ in the series (cf. \ref{sec:resum}). We have furthermore calculated the specific heat capacity for a larger cluster with $N=48$ sites and periodic boundary conditions using our  SU(2) symmetric finite temperature DMRG method with finite bond dimensions ranging from $\chi=2000$ to $\chi=10000$. These results were obtained from a numerical derivative of the (spline interpolated) energy as a function of inverse temperature.
For $T>2$, the results are converged with bond dimension and agree with the NLCE and finite $N=32$ cluster results (bottom panel). At lower temperatures, the dependence of the results on the bond dimension becomes significant and we extrapolate to infinite bond dimensions using a quadratic polynomial in $1/\chi$ yielding a very good match with the $N=32$ and NLCE results within the accessible temperature range ($T>0.62$). The errorbars indicate the distance of the extrapolated value from the largest bond dimension ($\chi=10000$).

\subsection{Heat capacity in zero field}
\label{ssec:Cvzero}

The specific heat capacity quantifies the change of the internal energy as a function of temperature and is directly accessible in experiment. Our results and a comparison to experimental results are summarised in  Fig.~\ref{fig:specheat_exp}. 

Starting at high temperatures, $T\gg1$, the heat capacity decays as $1/T^2$, as required in the  leading order high temperature expansion. In the regime down to $T=2$, all orders of the NLCE with $n>5$ agree with each other, and also  with the finite size $N=32$ result from typicality. 
The Euler transform of our NLCE results agrees between orders $n=7$ and $n=8$ down to $T\approx 0.25$, which we take to indicate that the series is converged over this range. 

Crucially, this allows us to resolve unambiguously the maximum of the heat capacity located at $T\approx 0.57$. In the proximity of the maximum, the results for the finite size cluster $N=32$ deviate significantly from the NLCE, which indicates that in the regime $T<2$ the correlations beyond the size of the $N=32$ cluster start playing a discernible role. 
From the strong dependence of the DMRG results on the bond dimension around the location of the maximum of the  heat capacity, as well as from the visible discrepancy of the specific heat for finite size clusters compared to the converged NLCE results close to the maximum, we conclude that the system enters a nontrivial quantum regime at temperatures $T\approx1$, where it exhibits entanglement beyond what is representable faithfully by $\chi=10000$ matrix product states.

\begin{figure}[h]
    \centering
    \includegraphics{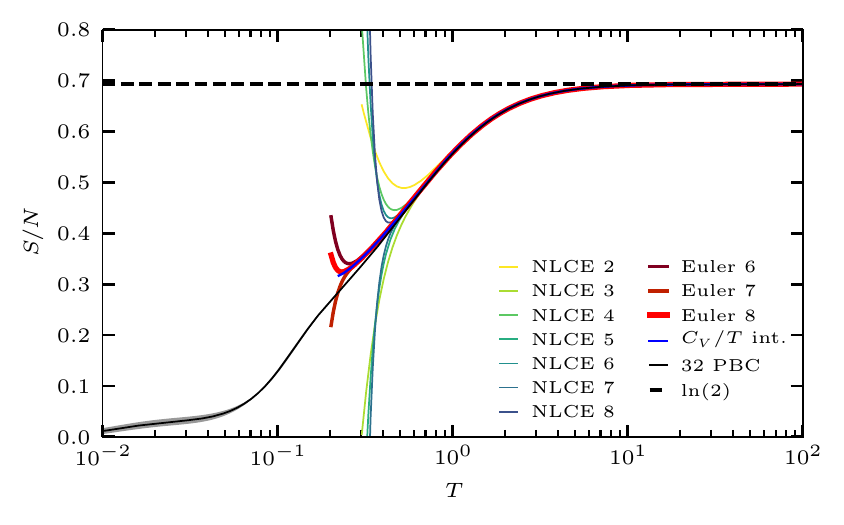}
    \caption{NLCE results for the thermodynamic entropy per lattice site $S/N$ as a function of temperature $T$. The yellow through dark blue curves show raw NLCE data for different expansion order $n$ up to $n=8$ tetrahedra. The brown, black and red curve are the corresponding Euler transform, showing converged entropies down to $T\approx0.25$ for $n=8$. These results agree  with the temperature integral (light blue curve) of $C_V/T$ of the Euler $n=8$ data for the specific heat from Fig. \ref{fig:spec_heat_synopsis}. We also show the entropy obtained from the finite size $32$ site cluster obtained from canonical typicality in the black curve.}
    \label{fig:entropy}
\end{figure}

The heat capacity of the $N=32$ cluster exhibits a second maximum at low temperatures, similarly to what was observed previously in a different pyrochlore model \cite{changlani_quantum_2018}. However due to the divergence from converged results of the NLCE in this regime, which is not subject to finite-size effects of this kind, we conclude that this feature is likely not representative for the thermodynamic limit.

Indeed, the converged part of our NLCE data shows a rapid decrease of the heat
capacity as $T$ is lowered from the maximum. Therefore, 
if there is an additional feature, it must be well separated from the maximum we have found.

In order to gain further insight into the low-$T$ regime, we have 
calculated the thermodynamic entropy as a function of temperature, 
\begin{equation}
    S(T_2) - S(T_1) = \int \limits_{T_1}^{T_2}\ \D T \frac{C_V}{T}\ ,
    \label{eq:entropy_spec_heat}
\end{equation}
with $S(\infty)=\ln 2$.  
A direct calculation of the entropy per site in NLCE using Eq. \eqref{eq:entropy} shown in Fig. \ref{fig:entropy} indeed agrees  with this temperature integral down to the lowest temperatures for which our (Euler transformed) NLCE is converged.

Interestingly, we find that just over half the total entropy is released down to $T\approx 0.25$, where the entropy is $S/N\approx0.33\approx 0.47\ln 2$. This in turn means that the spectral weight below the maximum is huge, and there is
plenty of scope for further interesting behaviour, the nature of which we are unfortunately unable to determine from our 
approach. [In several classical models\cite{castelnovo_spin_2012}, as well as some fine-tuned quantum models \cite{klein_exact_1982,chayes_valence_1989,Raman_SU2_2005}, not all the entropy is released even at $T=0$, but in real systems, the third law of thermodynamics 
stipulates that this is not what actually happens. For instance, for the finite size $N=32$ cluster, a steep decrease at low $T\alt 0.1$ is associated with its low-$T$ peak in the specific heat.] 

We will return to more detailed comparisons of this behaviour with other models and experimental systems in the final discussion, Sec.~\ref{sec:conclusion}.

\subsection{Magnetic susceptibility at $h=0$}
\label{ssec:chizero}

\begin{figure}[h]
    \centering
    \includegraphics[width=\columnwidth]{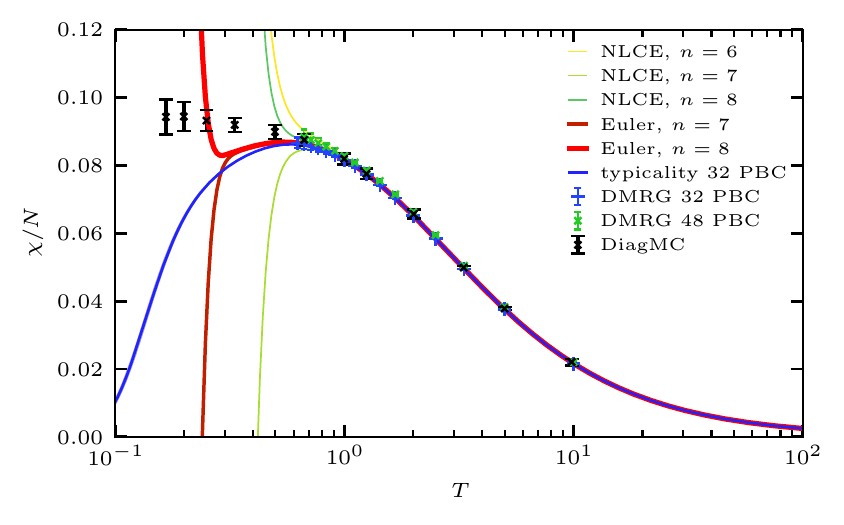}
    \caption{Magnetic susceptibility $\chi/N$  as a function of temperature $T$. We show raw data from different NLCE orders (yellow, green) as well as their Euler transform (red). The blue curve shows the result in the finite size cluster with $N=32$ spins and the black crosses with errorbars indicate the diagrammatic Monte Carlo results from Ref. \onlinecite{huang_spin-ice_2016}. We also show  DMRG results for the $N=32$ site cluster (blue crosses) and for the $N=48$ site cluster (green crosses) extrapolated to infinite bond dimension, where the errorbars indicate the distance of the extrapolated value to the largest bond dimension $\chi=10000$.
    }
    \label{fig:susceptibility_nlce}
\end{figure}

We consider the magnetic susceptibility $\chi/N$\footnote{We hope that there will be no confusion with the bond dimension in DMRG, for which we have conventionally used the same symbol.} per lattice site in Fig. \ref{fig:susceptibility_nlce} as a function of temperature. 
As in the case of the heat capacity, we perform NLCE calculations up to clusters of 8 tetrahedra and apply the Euler transform to these results. Fig. \ref{fig:susceptibility_nlce} shows the raw NLCE results at order $n=8$ to be converged down to temperatures of about $T\approx 0.8$. The Euler transform improves the convergence of the series significantly, again down to a temperature of $T\approx 0.25$.

At high temperatures $T\gtrsim 1$, the susceptibility obtained by the various methods agrees:
typicality for $N=32$ site cluster,  the $N=48$ cluster result obtained in finite temperature DMRG, extrapolated to infinite bond dimension using a quadratic polynomial in inverse bond dimension. 
At $T\approx 0.6$, the finite size magnetic susceptibility exhibits a pronounced maximum and decays rapidly to zero at lower temperatures. The Euler transform for the largest NLCE order clearly reveals a decrease of the magnetic susceptibility after a \emph{maximum} at $T=0.54$ in the thermodynamic limit.

We note that the  magnetic susceptibility for very large system sizes was previously calculated in diagrammatic Quantum Monte Carlo simulations in Ref. \onlinecite{huang_spin-ice_2016}, corresponding to the black crosses with errorbars in Figs. \ref{fig:susceptibility_nlce}. These results agree  with our NLCE and finite size results at temperatures above the maximum of the susceptibility. At low temperatures, however, they suggest a steady increase or plateau of the susceptibility instead of the  maximum that we find. It would be desirable to push both our and the diagrammatic Monte Carlo method to higher orders in order to see which of the two apparently irreconcilable behaviours is the correct one.   

\subsection{Static spin structure factor at zero field}
\label{ssec:sqzero}

The static spin structure factor quantifies the spin correlation patterns present at a given $T$: 

\[
S(\vec{Q})=\frac{4}{3N} \sum_{ij} \langle \vec{S}_i \cdot \vec{S}_j \rangle_\beta \cos \left[ \vec{Q} \cdot \left( \vec{R}_i - \vec{R}_j \right) \right] \ .
\]
\begin{figure}[h]
    \centering
        \includegraphics[width=\columnwidth]{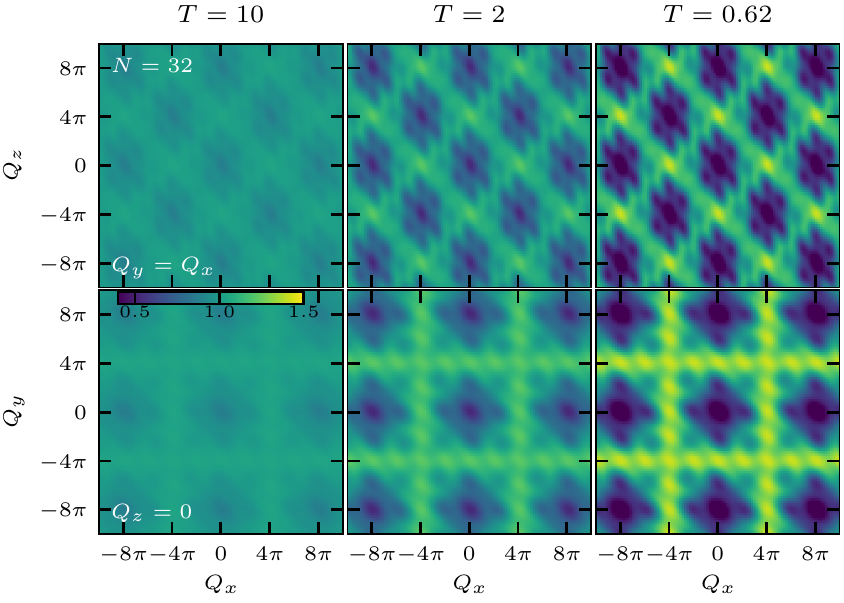}\\
           \includegraphics[width=\columnwidth]{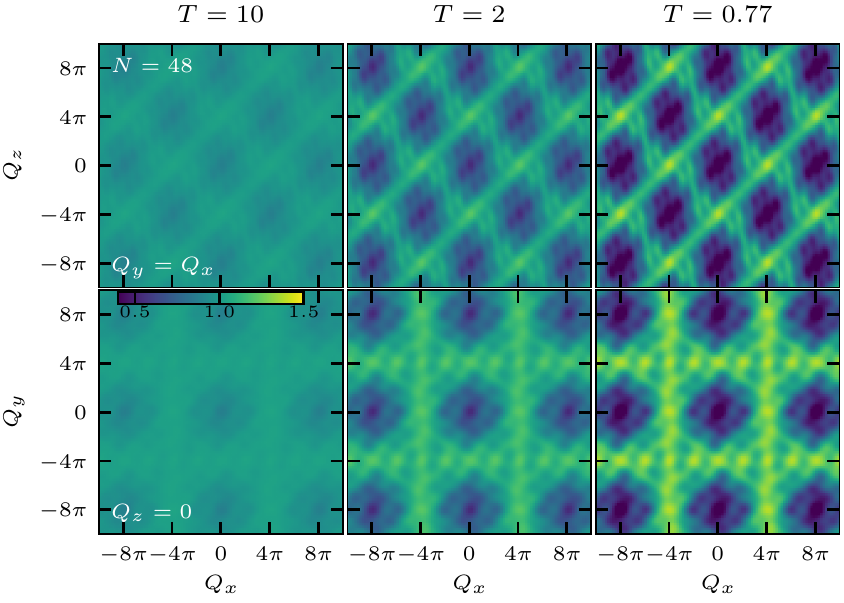}
    \caption{Static spin structure factor $S(\vec{Q})$ for different temperatures calculated with finite-$T$ DMRG in the $N=32$, upper rows ($N=48$, lower rows) site cluster using bond dimension of $\chi=8000$ ($\chi=10000$). Even rows show the $(h,h,l)$ plane in momentum space, odd  rows  the $(h,l,0)$ plane. The DMRG ansatz for the two different clusters has a different symmetry.}
    \label{fig:struct48}
\end{figure}

Here, we use finite temperature DMRG on the two clusters with $32$ sites and full cubic symmetry and $48$ sites with reduced symmetry, to investigate the static spin structure factor at finite temperature. Fig. \ref{fig:struct48} shows the result in the $32$ site cluster (top rows) and
 for the $48$ site cluster (bottom rows), in the  the $(h,h,l)$ plane (i.e. $Q_x=Q_y$, rows 1, 3) and the $(h,l,0)$ plane ($Q_z=0$, rows 2, 4) of  momentum space, extended over several Brillouin zones for different temperatures (columns). 
Both figures show the clear emergence of a correlation structure already at high $T$. This becomes more pronounced
as $T$ is lowered, without acquiring much additional structure: certainly, as expected
for a highly frustrated magnet, no sharp Bragg 
peaks appear, which would have been indicative of magnetic ordering. 

Indeed, with decreasing temperature, the weight in the center of the Brillouin zone ($\vec{Q}=0$) decreases and moves to the boundary of the extended Brillouin zone, the most visible location of increasing intensity being at $(0,0,\pm 4\pi)$. 

At this location, alongside $(\pm2\pi,\pm2\pi,\pm2\pi)$, one finds the incipient 
pinch points, well-known from other pyrochlore 
magnets \cite{harris_ordering_1998,CanalsLacroix_prl,Isakov_dipolar_prl,iqbal_quantum_2019,muller_thermodynamics_2019,plumb_continuum_2019}, as well as settings with an emergent U(1) gauge field more generally.

\begin{figure}[h]
    \centering
    \includegraphics[width=\columnwidth]{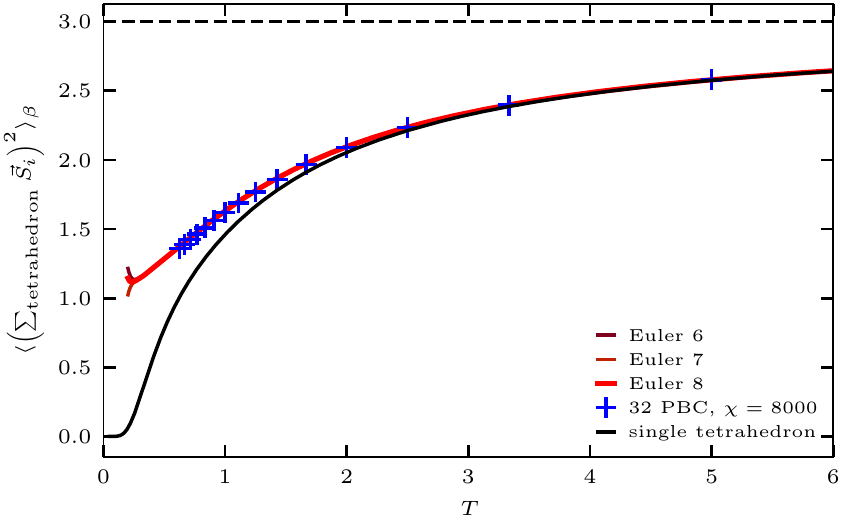}
    \caption{Total spin $\langle \sum_\text{tet} \vec{S}_i \rangle_\beta$ of a tetrahedron in the 32 (blue) site cluster as a function of temperature $T$. The red curves show the Euler transform of NLCE results and the black curve is the total spin of a single tetrahedron ($4$ site cluster) for comparison. }
    \label{fig:totspin}
\end{figure}

Due to the low resolution in $k$-space our finite system sizes up to $N=48$ do not permit to investigate more closely how sharp the pinch points become, since the associated  lengths at low temperatures may become longer than our linear cluster sizes. However, a {\it very} rough idea can be gleaned by considering the $T$-dependence of the energy. The reason for this is that the energy simply encodes the total spin of each tetrahedron, 
\begin{equation}
E=\sum_\mathrm{tet}\frac12\left(\sum_{i\in\mathrm{tet}} \vec{S}_i\right)^2 + \mathrm{const}.
\end{equation}
The pinch-points become infinitely sharp in the limit of vanishing tetrahedral magnetic 
moment, $\sum_{i\in\mathrm{tet}} \vec{S}_i=0$.\cite{Isakov_dipolar_prl} 

This condition is known to be met in classical Ising (where it is known as the ice rule\cite{anderson_ferrite,harris_geometrical_1997,castelnovo_spin_2012}), XY and Heisenberg models,  but it cannot
hold in the quantum realm, since a spin which is part of two tetrahedra cannot 
enter singlet bonds in both of them simultaneously. The deviation of the  
energy from the minmal value for  is thence a proxy
for the possible sharpness of the pinch-points. The total spin of a tetrahedron is plotted in Fig.~\ref{fig:totspin} from our
various methods. Eyeballing an extrapolation of this curve to $T=0$, one finds   
$\sum_{i\in\mathrm{tet}} \vec{S}_i\approx 1$, which confirms that frustration precludes that all tetrahedra are in singlet states and hence a finite width of the pinch points; for an estimate of the pinch-point width based on PF-FRG, see Ref.~\onlinecite{iqbal_quantum_2019}.

\subsection{Heat capacity and entropy in a magnetic field}
\label{ssec:Cvfinite}

We complement the above discussion by a further analysis of the behavior of the pyrochlore Heisenberg magnet in the presence of a finite field. For readability of our figures, we only show the converged part of eighth order NLCE results using the tetrahedra expansion. As before, we use the agreement of eighth and seventh order Euler transform as convergence criterion. 
\begin{figure}[h]
    \centering
    \includegraphics{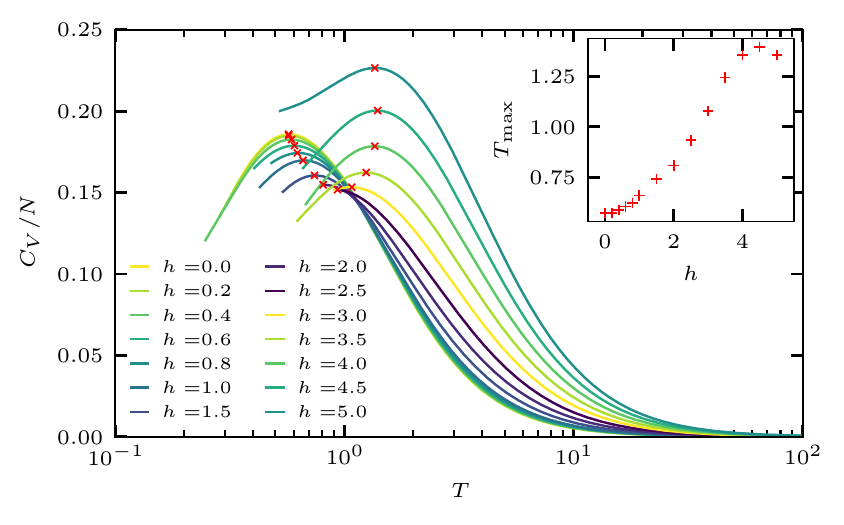}\\
       \includegraphics{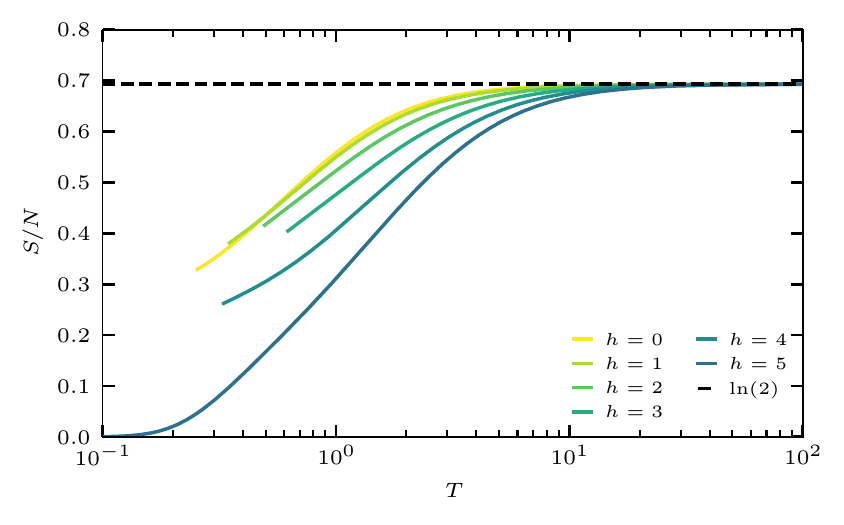}
    \caption{Magnetothermodynamics of the $S=1/2$ pyrochlore Heisenberg magnet. Top: heat capacity $C_V/N$ per lattice site at varying  fields as a function of temperature. We show data obtained from the Euler transform of the $n=8$ tetrahedra from NLCE. The shown temperature range corresponds to the part of the curve where the $n=7$ Euler transform agrees with the $n=8$ result to ensure convergence. Red crosses indicate the position of the maximum of the specific heat and the inset shows the temperature of the maximum as a function of the applied field $h$. Bottom: Entropy per site $S/N$ at varying fields as a function of temperature. We show only the converged part of the $n=8$ Euler tranform of our NLCE results.}
    \label{fig:cv_h}
       \label{fig:entropy_h}
\end{figure}

Fig. \ref{fig:cv_h} shows the heat capacity and entropy per site as a function of temperature for a range of different fields $h$ applied in the $[001]$ direction. We observe a shift of the maximum of the specific heat to higher temperatures at strong fields, as well as a non-monotonic change of the height of the maximum.

An overall upward shift of the weight is not particularly surprising given the presence of an additional term in the Hamiltonian. Indeed, at high fields, the curve resembles an unspectacular paramagnet. 

However, the structure of the low-energy spectral weight and its rearrangement at intermediate fields is complex. The weight at lowest energies is in large part  due to non-magnetic states which are not favoured by the magnetic field. At the same time, the more numerous states with nonzero magnetisation spread out as the field is applied. The failure of the Euler transform to converge to similarly low temperatures at intermediate fields is presumably due to a more complex behaviour of the specific heat in this intermediate field regime.  

\subsection{Magnetization at $h>0$}

\begin{figure}[h]
    \centering
    \includegraphics[width=\columnwidth]{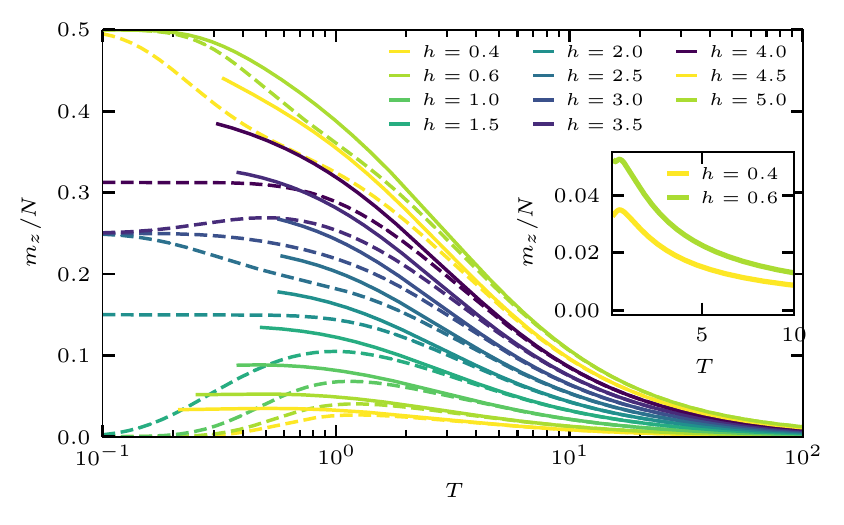}
    \caption{Magnetization $m_z/N$ as a function of $T$. We show results obtained from the $n=8$ Euler transform of the NLCE including data from clusters of up to $n=8$ tetrahedra, in the range where they agree with the $n=7$ Euler transform. At low fields $h=0.4$, $h=0.8$, we observe a local maximum of the finite temperature magnetization (inset). The dashed lines correspond to the magnetization per site for a single tetrahedron. }
    \label{fig:mag}
\end{figure}

We finally consider the effect of a magnetic field on the magnetization per site $m_z/N$. At zero field, there is no net magnetization in the absence of spontaneous symmetry breaking. Fig. \ref{fig:mag} shows converged NLCE results (with the highest order Euler transform) for the finite temperature magnetization (solid lines) in comparison with the result for a single tetrahedron (dashed lines). At high temperatures, these results agree and yield a Curie law. At intermediate temperatures, a difference due to the finite size of the tetrahedron is noticable, with the selection of different total magnetization groundstates in the low-$T$ limit, with $m_z/N$ being $0$, $0.25$, $0.5$ respectively depending on the field \footnote{Note that accidental degeneracies of the groundstate across different $S_z$ sectors lead to intermediate values of the magnetization of the single tetrahedron at $h=2$ and $h=4$.}. 

Interestingly, a nonmonotonic dependence of the magnetisation on $T$ can be observed at low fields, Fig.~\ref{fig:mag}, both in the NLCE results (emphasized in the inset)
as well as -- more visibly --  in the single tetrahedron case: the magnetisation vanishes
at both low $T$, because the weak-field low-energy states are dominantly non-magnetic; and at
high $T$, for the usual entropic reasons. At intermediate $T$, by contrast, the large weight
of magnetic states oriented by the fields dominates, whence the maximum.  

\section{Discussion and experiment}
\label{sec:conclusion}
Having laid out the results, we now place them in the broader context of other highly frustrated model systems on one hand, and experiments on magnetic materials on the other. Here, we focus
on the specific heat, not only because it is a quantity which is readily available across the board; but also because it allows for a relatively straightforward comparison between different platforms 
thanks to the integral $\frac1{N}\int_0^\infty dT^\prime \, C_V(T^\prime)/T^\prime=k_B\ln(2S+1)$.

We have collated the data for a number of models and materials in Fig.~\ref{fig:specheat_exp}. There, we compare (i) our converged results with what is found for single tetrahedra with (ii) $S=1/2$ or (iii) in the classical limit, $S=\infty$; as well as experiments on (iv) the Ising spin ice pyrochlore magnet Dy$_2$Ti$_2$O$_7$ and (v) the $S=1$ Heisenberg antiferromagnet NaCaNi$_2$F$_7$, rescaled by $\ln2/\ln3$ to take into account the greater spin length.
The $T$-axes of the experimental data, (iv,v), have been scaled so that the temperature of their respective maxima coincide with the one of our data; while the single tetrahedron results (ii,iii) were scaled to agree in the asymptotic limit of high-$T$. 

All of these have in common a considerable spectral weight downshift -- at $T=0.25$, all of them
exhibit a significant residual entropy, see inset of Fig.~\ref{fig:specheat_exp}. There are, however, considerable differences of detail (leaving aside case (iii) on account of the unbounded 
classical entropy). The single $S=1/2$ tetrahedron, (ii), 
releases its entropy more swiftly at low-$T$ than 
our NLCE results  on
account of its singlet gap coupled  with a small residual  entropy $S(0)=\frac14\ln2$. 
The spin
ice experiment, (iv), with a slightly higher residual entropy of around $S_p=\frac12\ln\frac32$, in fact
releases its entropy even more swiftly, with the peak in the specific heat peak being the most narrow
on the low-$T$ side. 

The NaCaNi$_2$F$_7$ experiment, (v), shows an initial high-$T$ release of the entropy remarkably close
to that of our $S=1/2$ results. However, already above the peak in $C_V$, the release in NaCaNi$_2$F$_7$ 
is
comparatively considerably greater, meaning that the spectral weight downshift in our 
results is stronger. 

Indeed, the breadth of the peak in $C_V$ we find for the $S=1/2$ Heisenberg model is
broader not only than all the cases (ii-v), but also than the other paradigmatic highly frustrated
 $S=1/2$ 
Heisenberg model, that on the kagome lattice. 
By comparison with the high order series expansion results in Ref.~\onlinecite{Misguich_2005}, we find that in the isotropic $S=1/2$ Heisenberg antiferromagnet on the kagome lattice, the residual entropy of $0.475 k_B \ln 2$ is reached already at a higher temperature of $T\approx 0.30$, whereas in the pyrochlore lattice our results suggest that this entropy is retained at a lower temperature of $T\approx 0.254$, corresponding to the larger spectral downshift in pyrochlore.

It should be emphasized that this is not at all what would obviously have been expected. 
Generally, low dimensionality is considered to favour spectral weight downshift, as encoded
e.g.~by the Mermin-Wagner theorem. Also, in the Ising setting, triangular motifs are considerably
more frustrated --- $S_\text{kagome} = \frac{1}{24 \pi^2} \int_0^{2\pi} \mathrm{d}x\mathrm{d}y \, \ln \left[21-4\left( \cos x + \cos y + \cos(x+y) \right) \right]  \approx0.50183$ \cite{kano_antiferromagnetism_1953} for the kagome Ising magnet, {\it much} larger than in the pyrochlore case
$S_\text{Pauling}\approx \frac 1 2 \ln \frac 3 2\approx 0.2027$\cite{pauling_structure_1935,anderson_ferrite}.

This implies that there is huge scope for unusual behaviour of this model at low-$T$. 
Alas, our results provide little indication of the detailed nature of the low-energy 
space of states. Indeed, many proposals for the behaviour of this magnet have been 
made, and it is hard to choose between them based on presently available information,
as there is not even compelling evidence in favour of a particular physical picture. 
The concurrent lack of a pristine experimental realisation goes a long way towards explaining
the divergence of theoretical predictions\cite{Raman_SU2_2005,iqbal_quantum_2019,CanalsLacroix_prl,reimers_mean-field_1991,MoessnerGoerbig_prb_2006,Berg_subcontractor_2003}
so that different methods arguably come up with the conclusion
most suited to them. 

We are therefore left with the twin higher-level insights, namely that the pyrochlore $S=1/2$ 
Heisenberg magnet is at least as frustrated, and arguably interesting, as the one on the kagome lattice; and that it is at least as intractable. We hope that future work will be able to build on 
the advances reported in this work. And, of course, that the low-$T$ regime will
become accessible experimentally in a suitable magnetic material. 

{\bf Note added:} As we were concluding this work, a preprint \onlinecite{derzhko_adapting_2020} appeared which also studied the thermodynamic properties of the pyrochlore $S=1/2$ Heisenberg model using a combination of methods including canonical typicality, high temperature series expansion and the entropy method. It placed particular emphasis on extrapolation schemes in order to access the  low-$T$ regime. 

\begin{acknowledgments}
We are very grateful to Kemp Plumb and Art Ramirez for kindly 
supplying the experimental data on NaCaNi$_2$F$_7$ and spin ice. 
We thank Owen Benton, Ludovic Jaubert, Paul McClarty, Jeffrey Rau, Johannes Richter, Oleg Derzhko, 
Masafumi Udagawa and Karlo Penc for very helpful discussions.
We acknowledge financial support from the Deutsche Forschungsgemeinschaft
through SFB 1143 (project-id 
247310070) and cluster of excellence ct.qmat (EXC 2147, project-id 39085490). 
I.H.~was supported in part by the Hungarian National Research,   
Development   and   Innovation Office (NKFIH) through Grant No.~K120569 and 
the Hungarian  Quantum  Technology  National  Excellence Program  (Project  
No.~2017-1.2.1-NKP-2017-00001). Some of the data presented here was produced 
using the \textsc{SyTen} toolkit, originally created by Claudius 
Hubig.\cite{hubig:_syten_toolk,hubig17:_symmet_protec_tensor_networ}
\end{acknowledgments}

\appendix
\newpage
\section{Comparison of NLCE expansions}
\label{sec:nlcecomp}

The NLCE expansion is general in that it pemits in principle a wealth of different cluster expansions based on constraints on the included class of clusters. 
Since the number of clusters grows factorially with the number of constituents, it is often wise to constrain the class sufficiently in order to get a managable number of clusters at the largest cluster sizes which are still solvable in full diagonalization.

\begin{figure}[h]
    \centering
    \includegraphics[width=\columnwidth]{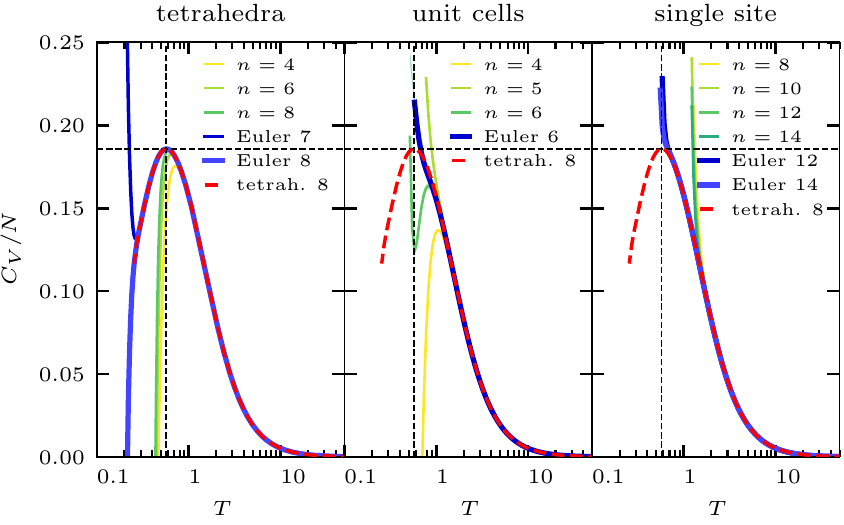}
    \caption{Convergence comparison of different NLCE expansions. Left: Expansion using clusters with complete tetrahedra up to 8th order. Center: Expansion using complete unit cells up to 6th order (100 clusters with $n=6$ unit cells: included 48 clusters completely with ED and 8 clusters in combination with canonical typicality). Right: Expansion using single sites, up to 14th order. For all expansions the highest order Euler series acceleration is shown. The expansion based on complete tetrahedra consistently yields superior convergence and is used in this work. Dashed lines show the height and position of the Schottky anomly as extracted from the left panel for comparison. The red dashed lines correspond to the 8th order Euler transform of the thetrahedra expansion, showing that all expansions agree with this result in the converged regime.}
    \label{fig:nlce_comp}
\end{figure}

\begin{figure}[h]
    \centering
    \includegraphics[width=\columnwidth]{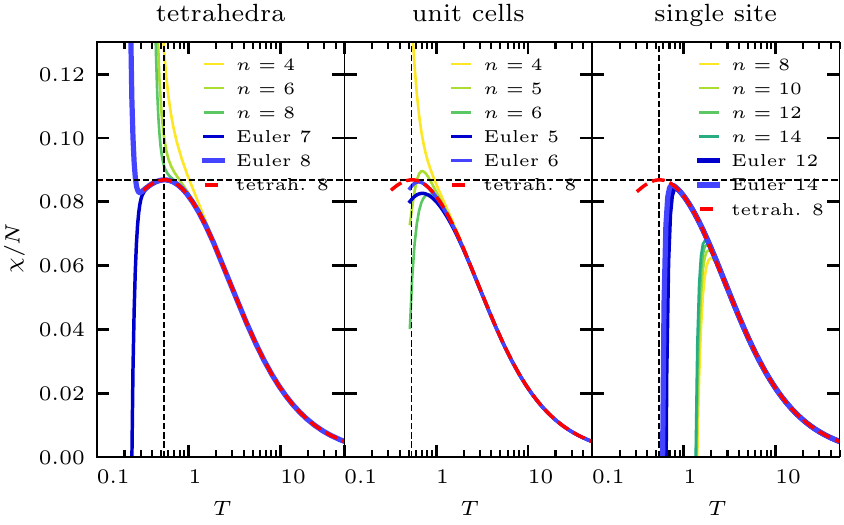}
    \caption{Convergence comparison of different NLCE expansions for the susceptibility. Left: Expansion using clusters with complete tetrahedra up to 8th order. Center: Expansion using complete unit cells up to 6th order (100 clusters with $n=6$ unit cells: included 48 clusters completely with ED). Right: Expansion using single sites, up to 14th order. For all expansions the highest order Euler series acceleration is shown. The expansion based on complete tetrahedra consistently yields superior convergence and is used in the remainder of this work. Dashed lines show the height and position of the Schottky anomaly as extracted from the left panel for comparison.}
    \label{fig:nlce_comp_susc}
\end{figure}

These constraints should respect the underlying physical properties of the system as much as possible to get a rapidly converging series expansion. We show a comparison of three different NLCE expansions for the pyrochlore lattice in Figs. \ref{fig:nlce_comp} and \ref{fig:nlce_comp_susc}: the single site expansion, the unit cell expansion (i.e. clusters on the fcc lattice, decorated by the tetrahedral unit cell) and the tetrahedral expansion used in the main text.

All expansions converge to the same curve at high temperatures, however the highest order single site expansion does not reach temperatures low enough to resolve the maximum of the specific heat, even after the Euler transform is applied. The unit cell expansion includes in principle much larger clusters (up to 6 unit cells correspond to 24 site clusters), however it is barely possible to obtained converged results for lower temperatures than in the single site expansion, while the tetrahedral expansion yields convergence to much lower temperatures and a clear resolution of the specific heat capacity maximum.

There are several reasons for this superior behavior: First, the central motif of the pyrochlore lattice is the tetrahedron. It is crucial to avoid dangling bonds and incomplete tetrahedra, which are present in both the unit cell and single site expansion. Secondly, due to the construction of unit cell clusters, the clusters used here do not include similarly large loops as in the tetrahedral expansion, which are crucial at low temperatures. Thirdly, and this is purely technical, due to unfavorable symmetry properties of the $n=6$ unit cell clusters, we were unable to solve all 100 topologically distinct clusters with full diagonalization, since the largest remaining symmetry blocks were too large. Finally, the Euler transform for the unit cell expansion can only rely on 5 complete expansion orders, which is far less than in the case of the single site and tetrahedral expansion.

Therefore, we conclude that the tetrahedral expansion is far superior to any other approach in the pyrochlore lattice and allows to access the lowest temperatures.

\section{NLCE Clusters}\label{app:nlce}

\begin{figure}[h]
	\centering
    \includegraphics[width=0.8\columnwidth]{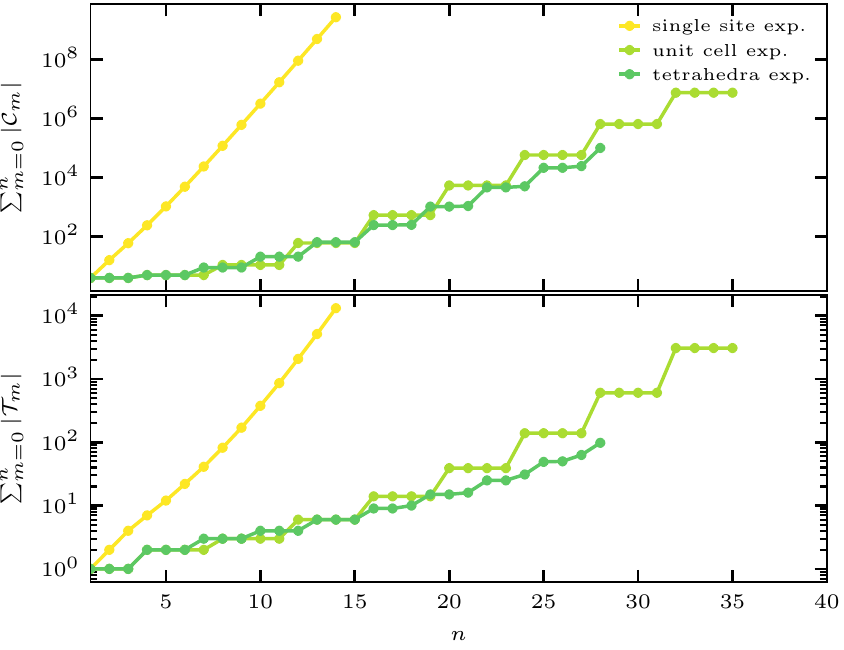}
	\caption{Number of connected $\vert\mathcal{C}_n\vert$ and topologically distinct clusters $\vert\mathcal{T}_n\vert$ for the three different expansions.}
	\label{fig:nlce_00}
\end{figure}

As mentioned in the previous section, a successful NLCE scheme needs to limit the growth of the number of clusters to an extent that the number of largest tractable clusters (using full diagonalization) is not too large. 

\begin{table}
	\begin{tabular}{| c | c | c | c |}
		\hline\hline
		$n$ & $\vert\mathcal{C}_n\vert$ &  $\vert\mathcal{S}_n\vert$ & $\vert\mathcal{T}_n\vert$\\\hline\hline 
		1 & 4 & 2 & 1 \\\hline
		2 & 12 & 2 & 1 \\\hline
		3 & 44 & 8 & 2 \\\hline
		4 & 182 & 19 & 3 \\\hline
		5 & 816 & 84 & 5 \\\hline
		6 & 3856 & 338 & 10 \\\hline
		7 & 18916 & 1650 & 19 \\\hline
		8 & 95436 & 8026 & 41 \\\hline
		9 & 492124 & 41370 & 88 \\\hline
		10 & 2582256 & 215564 & 207 \\\hline
		11 & 13743828 & 1147137 & 483 \\\hline
		12 & 74022676 & 6170524 & 1216 \\\hline
		13 & 402692008 & 33567270 & 3049 \\\hline
		14 & 2209562820 & 184140685 & 8002 \\\hline
	\end{tabular} 
	\caption{Number of connected $\vert\mathcal{C}_n\vert$, symmetrically distinct $\vert\mathcal{S}_n\vert$ and topologically distinct clusters $\vert\mathcal{T}_n\vert$ per unit cell listed with the order of expansion $n$ (equals system size) for the single site expansions.}
	\label{tab:app_nlce_00}
\end{table}

\begin{table}
	\begin{tabular}{| c | c | c | c | c |}
		\hline\hline
		$n$ & $N$ & $\vert\mathcal{C}_n\vert$ &  $\vert\mathcal{S}_n\vert$ & $\vert\mathcal{T}_n\vert$\\\hline\hline 
		0 & 1 & 4 & 2 & 1 \\\hline  		
		1 & 4 & 1 & 1 & 1 \\\hline
		2 & 8 & 6 & 2 & 1 \\\hline
		3 & 12 & 50 & 12 & 3 \\\hline
		4 & 16 & 475 & 90 & 8 \\\hline
		5 & 20 & 4881 & 844 & 25 \\\hline
		6 & 24 & 52835 & 8912 & 100 \\\hline
		7 & 28 & 593382 & 99252 & 466 \\\hline
		8 & 32 & 6849415 & 1142759 & 2473 \\\hline
	\end{tabular} 
	\caption{Number of connected $\vert\mathcal{C}_n\vert$, symmetrically distinct $\vert\mathcal{S}_n\vert$ and topologically distinct clusters $\vert\mathcal{T}_n\vert$ per unit cell listed with the order of expansion $n$ and system sizes $N$ for the unit cell expansions.}
	\label{tab:app_nlce_01}
\end{table}

\begin{table}
	\begin{tabular}{| c | c | c | c |}
		\hline\hline
		$n$ & $N$ & $\vert\mathcal{C}_n\vert$  & $\vert\mathcal{T}_n\vert$\\\hline\hline 
		0 & 1 & 4 & 1 \\\hline  		
		1 & 4 & 2 & 1 \\\hline
		2 & 7 & 4 & 1 \\\hline
		3 & 10 & 12 & 1 \\\hline
		4 & 13 & 44 & 2 \\\hline
		5 & 16 & 182 & 3 \\\hline
		6 & 18,19 & 796 & 6 \\\hline
		7 & 21,22 & 3612 & 10 \\\hline
		8 & 24,25 & 16786 & 24 \\\hline
		9 & 26,27,28 & 79426 & 49 \\\hline
	\end{tabular} 
	\caption{Number of connected $\vert\mathcal{C}_n\vert$, symmetrically distinct $\vert\mathcal{S}_n\vert$ and topologically distinct clusters $\vert\mathcal{T}_n\vert$ per unit cell listed with the order of expansion $n$ and system sizes $N$ for the tetrahedra expansions.}
	\label{tab:app_nlce_02}
\end{table}

In tables \ref{tab:app_nlce_00} through \ref{tab:app_nlce_02}, we list the number of clusters appearing at each order $n$ in the single site, unit cell and tetrahedral expansion. The order $n$ refers to the number of sites, unit cells, or tetrahedra respectively. The numbers listed correspond to the total number of clusters $|\mathcal{C}_n|$, the number of clusters which are not identical under application of non-translational symmetries $|\mathcal{S}_n|$ and the number of topologically distinct clusters $|\mathcal{T}_n|$, which is computationally relevant since this is the number of clusters for which the Hamiltonian has to be diagonalized. The growth of the number of clusters with order $n$ is depicted in Fig. \ref{fig:nlce_00}.

It should be noted that it is computationally challenging to check the topological equivalence of two clusters, since their interaction graphs need to be checked for an isomorphism, which is an NP hard problem. Therefore, e.g. the reduction of $184140685$ symmetrically distinct clusters at $n=14$ in the unit cell clusters to $8002$ topologically distinct clusters is already difficult and limits severely the access to higher orders.

\begin{figure*}[t]
    \centering
    \includegraphics{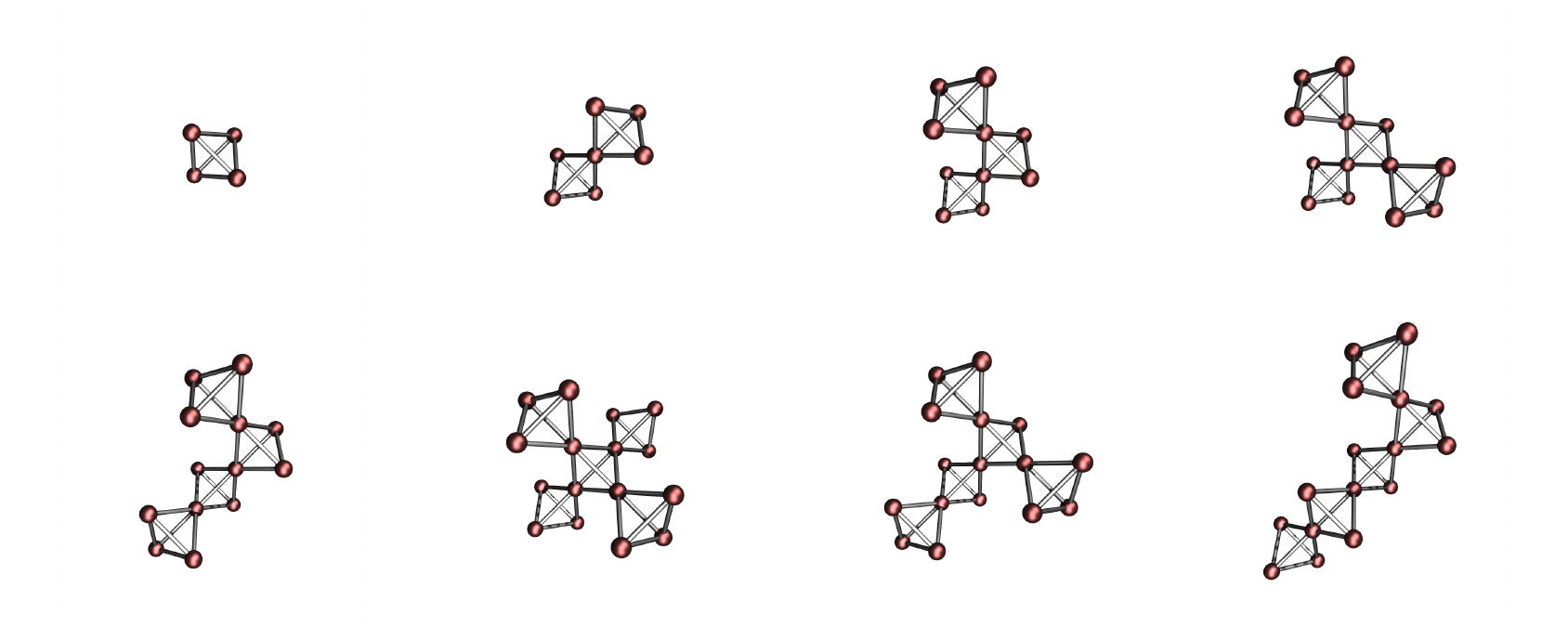}\\
    All eight topologically distinct clusters with one through five tetrahedra. 
    \centering
    \includegraphics{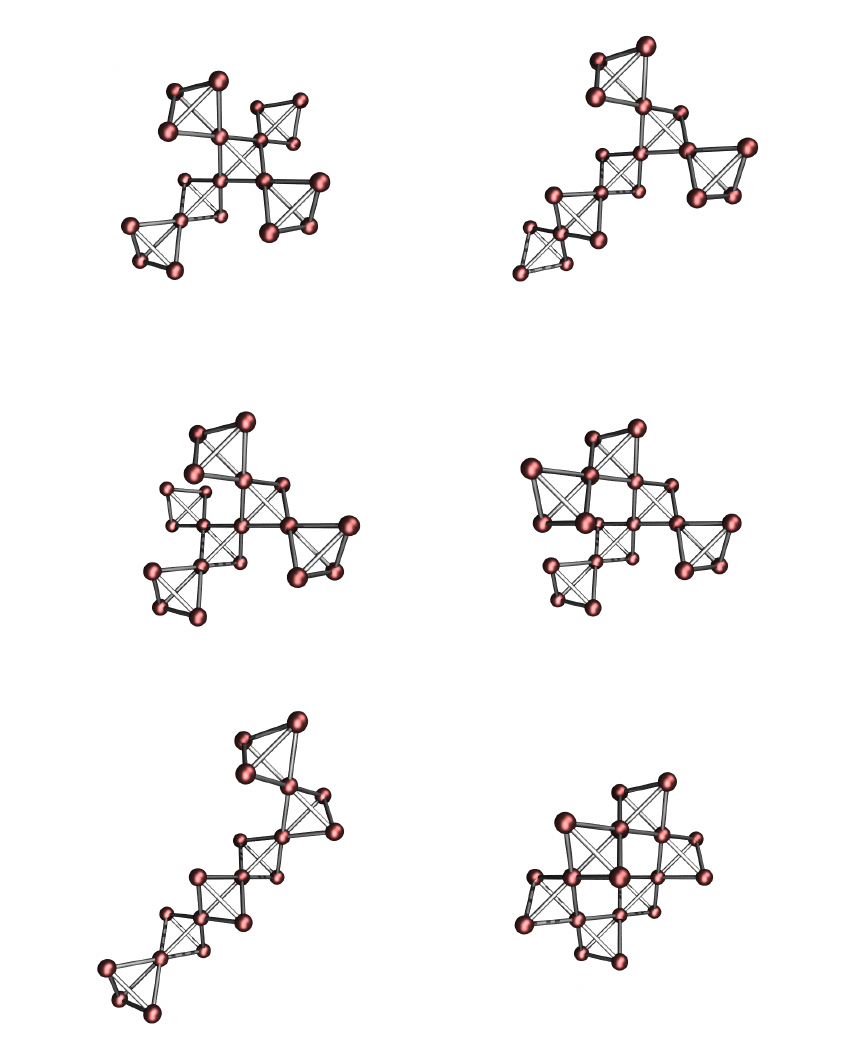}
    \caption{All 6 topologically distinct clusters with six tetrahedra. }
\label{fig:NLCE_6tetra}
\end{figure*}

\begin{figure*}[t]
    \centering
    \includegraphics{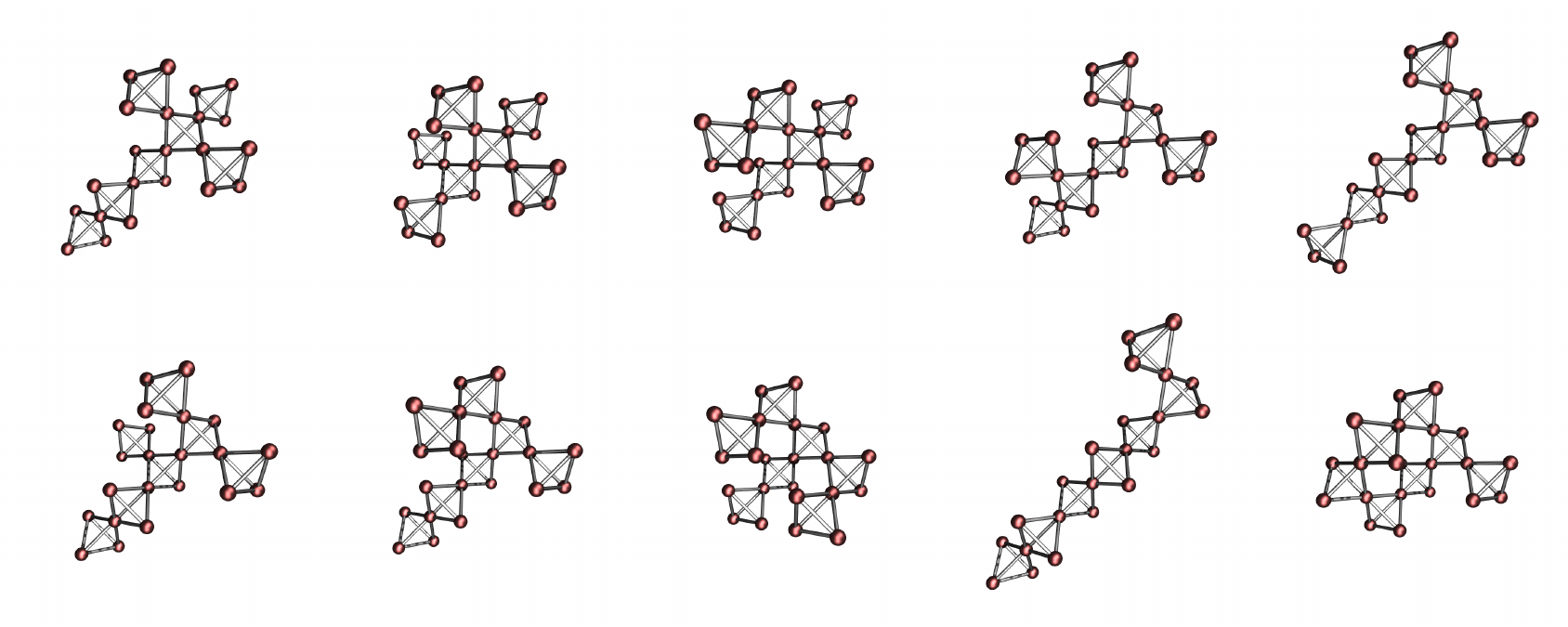}\\
    All 10 topologically distinct clusters with seven tetrahedra. 
    \includegraphics{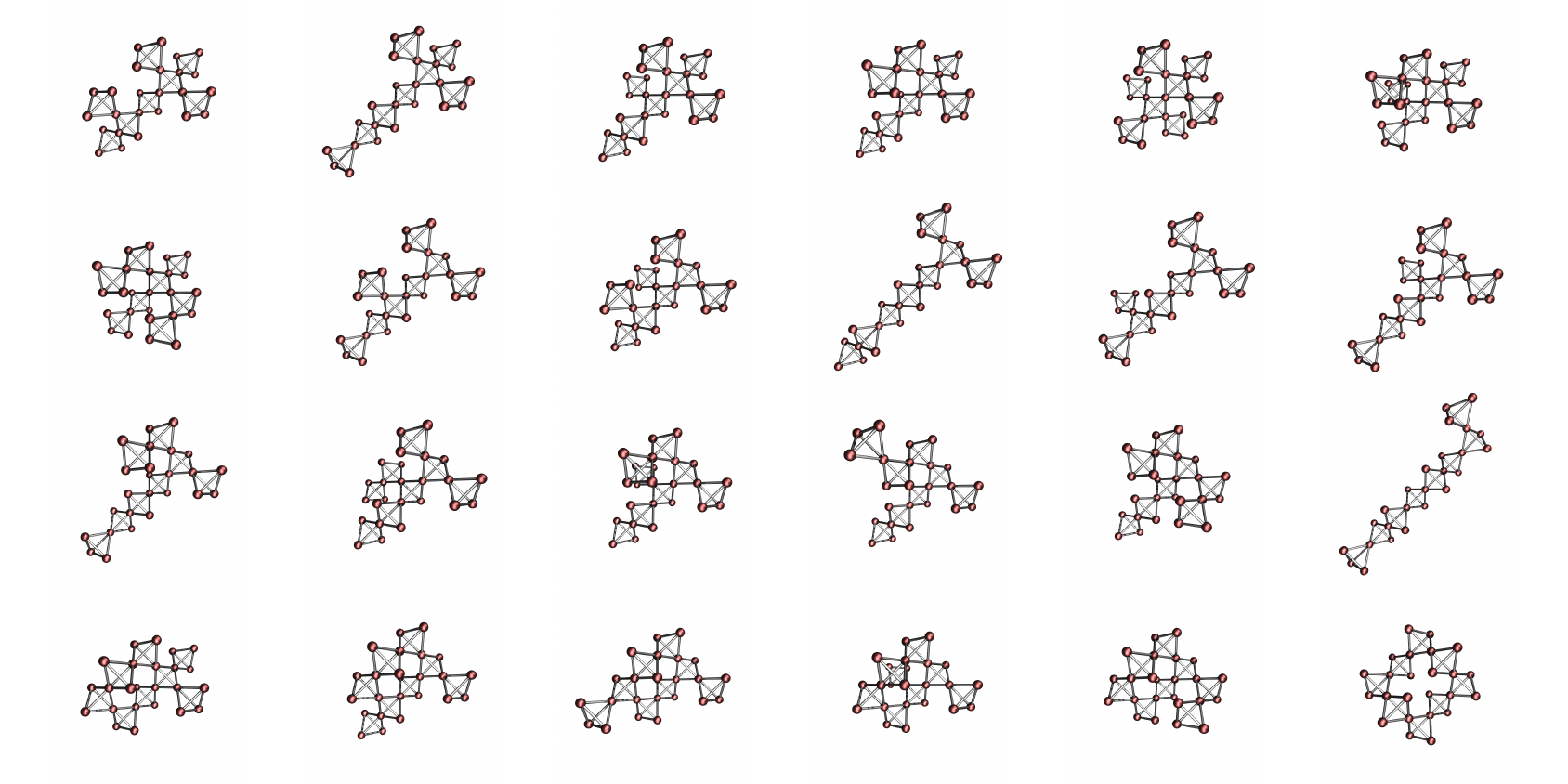}
    \caption{All 24 topologically distinct clusters with eight tetrahedra. }
\label{fig:NLCE_8tetra}
\end{figure*}

For the tetrahedral expansion, there is only one topologically distinct cluster for the orders 1 to 3 and two clusters with four tetrahedra. At the highest order we could reach, there are 24 clusters composed of eight tetrahedra, which have either 24 or 25 spins, just at the limit of what can be solved with full exact diagonalization using all symmetries of the clusters. We show all topologically distinct clusters included in the NLCE up to 8 tetrahedra in Figs. \ref{fig:NLCE_6tetra} and \ref{fig:NLCE_8tetra}.

\section{Finite size clusters}

\begin{figure}
	\includegraphics[width=\columnwidth]{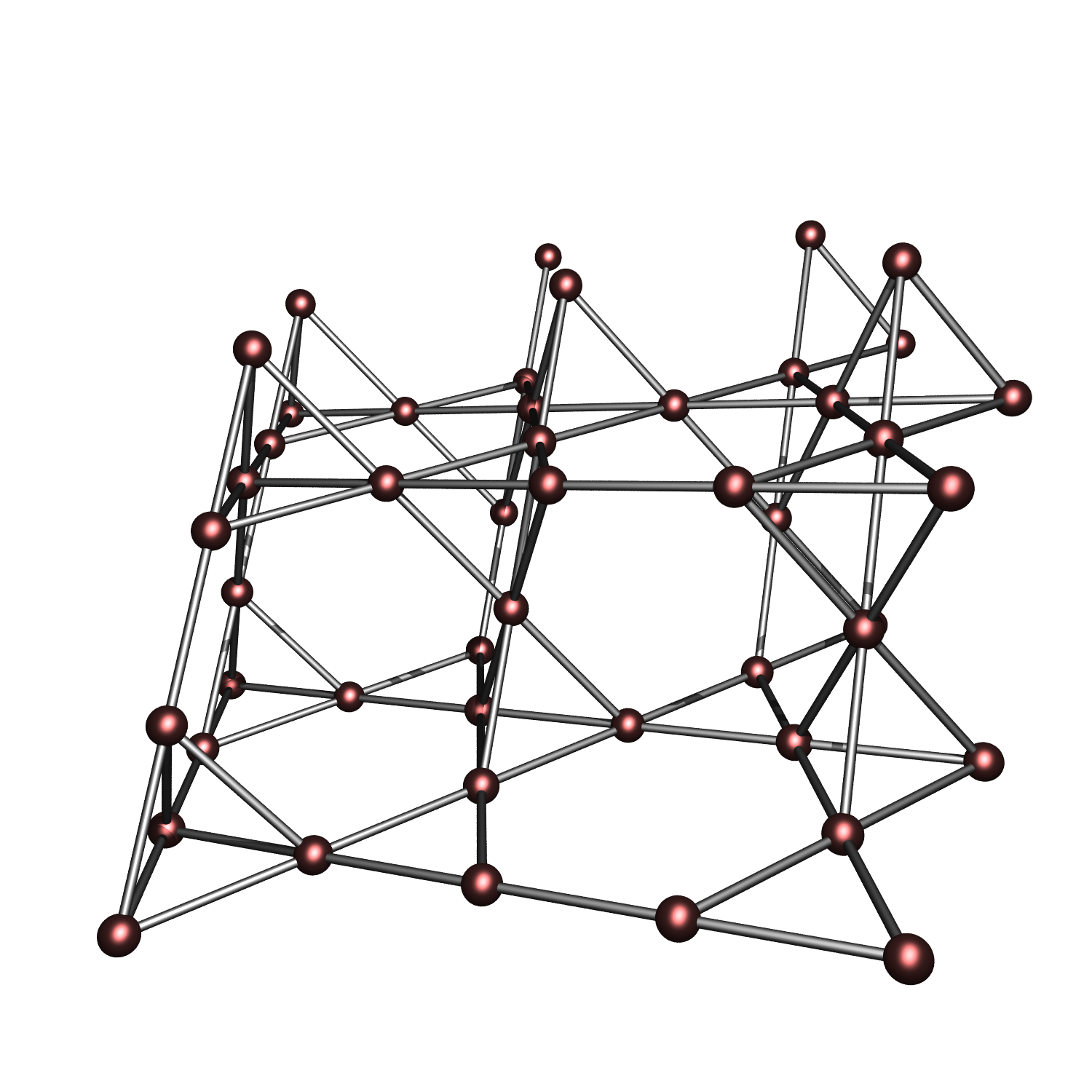}
	\caption{48 site cluster used in our finite temperature DMRG simulations. We use periodic boundary conditions (periodic bonds not shown for clarity).
	\label{fig:48site}
}
\end{figure}

In the present study, we consider two finite size clusters, the first being standard 32 site cluster consisting of two unit cells in direction $\vec{a}_1$, $\vec{a}_2$, $\vec{a}_3$ studied e.g. in Refs. \cite{changlani_quantum_2018,derzhko_adapting_2020}, which we depicted in the inset of Fig. \ref{fig:spec_heat_synopsis}. The second cluster we study is the 48 site cluster shown in Fig. \ref{fig:48site}

\clearpage
\bibliography{pyrochlore}
\end{document}